\documentclass[sn-nature]{sn-jnl}%

\usepackage{graphicx}%
\usepackage{multirow}%
\usepackage{amsmath,amssymb,amsfonts}%
\usepackage{bm}
\usepackage{mathrsfs}%
\usepackage[title]{appendix}%
\usepackage{xcolor}%
\usepackage{textcomp}%
\usepackage{manyfoot}%
\usepackage{booktabs}%
\usepackage[ruled, linesnumbered]{algorithm2e}
\usepackage{algorithmicx}%
\usepackage{algpseudocode}%
\usepackage{listings}%

\usepackage{braket}
\usepackage{bm}
\usepackage{comment}

\newcommand{\pdif}[2]{\frac{\partial #1}{\partial #2}}


\begin{document}

\title[Efficient computational homogenization]{Efficient computational homogenization via tensor train format}

\author*[1]{Yuki Sato}\email{yuki-sato@mosk.tytlabs.co.jp}

\author[1]{Yuto Lewis Terashima}

\author[1]{Ruho Kondo}

\affil*[1]{Toyota Central R\&D Labs., Inc., 1-4-14, Koraku, Bunkyo-ku, Tokyo, 112-0004, Japan}

\abstract{
Real-world physical systems, like composite materials and porous media, exhibit complex heterogeneities and multiscale nature, posing significant computational challenges. 
Computational homogenization is useful for predicting macroscopic properties from the microscopic material constitution. 
It involves defining a representative volume element (RVE), solving governing equations, and evaluating its properties such as conductivity and elasticity. 
Despite its effectiveness, the approach can be computationally expensive. 
This study proposes a tensor-train (TT)-based asymptotic homogenization method to address these challenges. By deriving boundary value problems at the microscale and expressing them in the TT format, the proposed method estimates material properties efficiently. 
We demonstrate its validity and effectiveness through numerical experiments applying the proposed method for homogenization of thermal conductivity and elasticity in two- and three-dimensional materials, offering a promising solution for handling the multiscale nature of heterogeneous systems.
}


\maketitle

\section*{Introduction}\label{sec:intro}
Real-world physical systems are inherently complex due to pronounced heterogeneities. 
Such heterogeneities, including composite materials and porous media, are characterized by their length scales, which present the multiscale nature of heterogeneous systems.
Although numerical simulation techniques serve to comprehend and predict the behavior of such systems, the detailed resolution of material constitution is often beyond current computational capabilities, which results in significant challenges in material science and engineering.

Computational homogenization~\cite{terada2000simulation, geers2010multi, geers2017homogenization, yvonnet2019computational} is a technique mainly used in material science and engineering to numerically predict the macroscopic properties of heterogeneous materials based on their microscopic material constitution e.g. different phases, grains, or fibers.
It serves as a powerful tool for bridging the gap between different scales--from microscopic material constitution, i.e. microscale, to a macroscopic structure of a material, i.e. macroscale.
Computational homogenization procedures typically consist of (1) defining a representative volume element (RVE), (2) solving the governing equations (GEs) in the RVE, and (3) evaluating the macroscopic properties, such as conductivity, elasticity, and permeability, using the solutions of GEs.
The RVE is chosen to include full features of heterogeneities so that the macroscopic material can be represented by the periodic structure of the RVE.
Computational homogenization has been widely applied to various multiscale systems, including composite materials~\cite{feyel2000, raju2021review}, crystal~\cite{roters2010overview}, and porous media~\cite{khoei2021computational}.
The GEs in the microscale are usually described as partial differential equations, which are solved by the finite element method (FEM), though some studies dealt with molecular dynamics as well~\cite{mortazavi2013combined, terashima2024fine}.
Since homogenization assumes the periodicity of microscale features, its computational cost can still be expensive when the periodical scale is much larger than the minimum length scale of material features due to the requirement of high-resolution numerical analysis.

The tensor-train (TT) format~\cite{oseledets2009compact, oseledets2011tensor, holtz2012manifolds, holtz2012alternating}, one of the tensor decompositions, is a powerful tool to represent a high-dimensional array efficiently in computational science and can be used for efficient high-resolution numerical analysis. 
The TT format is also known as matrix product states (MPSs) and matrix product operators (MPOs)~\cite{vidal2003efficient, schollwock2011density}, which are components of tensor networks~\cite{orus2014practical, montangero2018introduction, orus2019tensor, ran2020tensor}, in quantum physics.
Oseledets and Dolgov showed that the TT format can solve linear systems of equations faster than conventional full linear system solvers derived from elliptic partial differential equations~\cite{oseledets2012solution}.
There are also several studies for solving PDEs in the TT format or using the MPSs, including phase field models~\cite{risthaus2022solving} and fluid flows~\cite{gourianov2022quantum, kiffner2023tensor}.
Particularly, Gourianov et al.~\cite{gourianov2022quantum} applied the TT format to simulating turbulence and revealed that the scale-locality of the turbulence, that is, the feature of eddies with a certain length scale mainly interacting with other eddies with a similar scale, is well-suited to the TT format, analogous to local interactions in quantum many-body systems.
This motivates us to use the TT format or MPSs for computational homogenization where the heterogeneities can have scale-locality even in the RVE.

In this study, we propose the tensor-train-based asymptotic homogenization method to examine the effectiveness of the use of the TT format in multiscale systems.
Asymptotic homogenization~\cite{chung2001asymptotic, engquist2008asymptotic, kalamkarov2009asymptotic, penta2017introduction} is a kind of computational homogenization which relies on asymptotic expansions of the field variables (e.g. displacement, temperature, etc.) in terms of an infinitesimal parameter representing the ratio of the microscale dimensions to the macroscale dimensions.
We derive the boundary value problems at the microscale, discretized using the finite difference method, and express them including the material configuration in the TT format.
We demonstrate the validity and the effectiveness of our proposed method by applying it to the homogenization of the thermal conductivity and the elasticity for both two- and three-dimensional materials.

\section*{Results}
\subsection*{Target systems and asymptotic homogenization}
\begin{figure}[t]
    \centering
    \includegraphics[width=\textwidth]{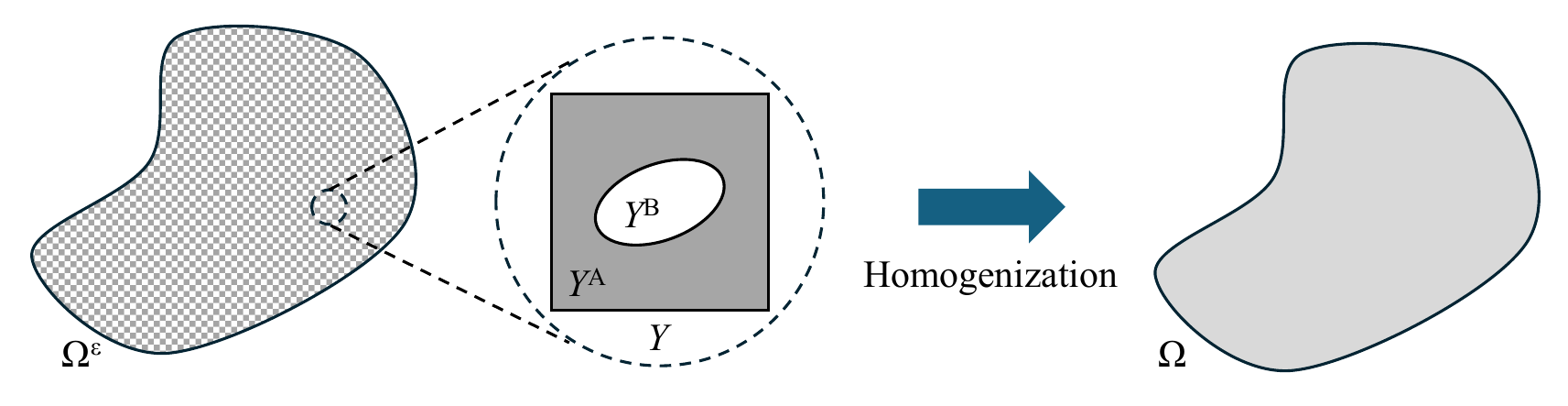}
    \caption{A conceptual diagram of computational homogenization.}
    \label{fig:homo}
\end{figure}

Consider a heterogeneous material domain $\Omega^\epsilon$ in the $d$ dimensional space consisting of materials A and B where $\epsilon$ is an infinitesimal parameter characterizing the length scale of heterogeneities.
Due to the heterogeneous property, the material constants depend on the spatial coordinate $\bm{x}$ in the material domain $\Omega^\epsilon$.
The homogenization techniques enable us to regard the heterogeneous material as a macroscopically equivalent homogeneous material $\Omega$.
Let $Y$ denote the RVE of the heterogeneous material where materials A and B occupy the domains $Y^\mathrm{A}$ and $Y^\mathrm{B} (= Y \setminus Y^\mathrm{A})$, respectively.
We also consider the homogeneous material $\Omega$ equivalent to the heterogeneous material $\Omega^\epsilon$ and introduce the characteristic length scales $L_\Omega$ and $L_Y$ to the homogeneous material $\Omega$ and the RVE $Y$, respectively.
We can characterize the heterogeneity of $\Omega^\epsilon$ by the ratio of the length scales $\epsilon := L_Y / L_\Omega$.
We illustrate the conceptual diagram of the homogenization in Fig.~\ref{fig:homo}.
By solving a characteristic equation defined on the RVE, we can estimate the homogenized material properties that characterize the macroscopic behavior of the material.
Specifically, let us consider the steady-state thermal conduction and the linear elasticity.

\paragraph{Thermal conductivity}
Let $\kappa^\mathrm{A}$ and $\kappa^\mathrm{B}$ denote the thermal conductivities of the isotropic materials A and B, respectively, and $\kappa(\bm{x})$ denote the spatially varying thermal conductivity which takes $\kappa^\mathrm{A}$ or $\kappa^\mathrm{B}$ depending on the coordinate $\bm{x}$.
The steady-state heat equation governing the heterogeneous material $\Omega^\epsilon$ is given as
\begin{align}
    \pdif{}{x_i} \left( \kappa(\bm{x}) \pdif{T^\epsilon (\bm{x}) }{x_i}  \right) = Q^\epsilon (\bm{x}) \text{ in } \Omega^\epsilon, \label{eq:heat_equation}
\end{align}
where $T^\epsilon$ is the temperature, $Q^\epsilon$ is the heat source, and $x_i$ is the $i$-th component of the coordinate $\bm{x}$ at an orthonormal basis $\{ \bm{e}_j \}_{j=0}^{d-1}$.
The asymptotic homogenization technique enables us to estimate the effective thermal conductivity tensor $\overline{\bm{\kappa}}$ of a heterogeneous material, as follows~\cite{engquist2008asymptotic, kalamkarov2009asymptotic}:
\begin{equation}
    \overline{\kappa}_{ij} = \frac{1}{|Y|} \int_Y \kappa(\bm{y}) \left( \delta_{ij} - \pdif{\phi^{j} (\bm{y}) }{y_{i}}  \right) \mathrm{d} \bm{y},
    \label{eq:homo_thermal}
\end{equation}
where $\overline{\kappa}_{ij}$ is the $(i,j)$-component of the thermal conductivity tensor $\overline{\bm{\kappa}}$ at an orthonormal basis $\{ \bm{e}_j \}_{j=0}^{d-1}$, and $\{ \phi^{j}(\bm{y}) \}_{j=0}^{d-1}$ is the characteristic temperature field satisfying
\begin{align} \label{eq:cp_thermal}
    \begin{cases}
        &\pdif{}{y_i} \left( \kappa(\bm{y}) \pdif{\phi^{j} (\bm{y})}{y_i} \right) = \pdif{ \kappa(\bm{y}) }{y_{j}} \text{ in } Y \\
        & \phi^{j} : Y \text{-periodic},
    \end{cases}
\end{align}
for $j \in 0, \dots d-1$.
The characteristic temperature field $\phi^{j}$ represents the temperature field induced by a unit macroscopic heat flux along with the $y_j$-axis.
The asymptotic homogenization technique consists of two steps: we first solve Eq.~\eqref{eq:cp_thermal} in the RVE to obtain the characteristic temperature field $\{ \phi^{j}(\bm{y}) \}_{j=0}^{d-1}$, and then calculate the homogenized thermal conductivity tensor by Eq.~\eqref{eq:homo_thermal}.

\paragraph{Linear elastic material}
We consider the static equilibrium of a linear elastic material.
The static equilibrium of the Cauchy stress tensor $\bm{\sigma}^\epsilon (\bm{x})$ and the external force vector $\bm{f}^\epsilon$ derives the governing equation as
\begin{align}
    \pdif{\sigma_{ji}^\epsilon (\bm{x})}{x_j} + f_i^\epsilon (\bm{x}) = 0, \label{eq:stress_equilibrium}
\end{align}
where $\sigma_{ji}^\epsilon$ and $f_i^\epsilon$ are the components of the Cauchy stress and the external force, respectively.
Let us assume that materials A and B are isotropic linear elastic materials with the Lam\'{e}'s constants $(\lambda^\mathrm{A}, \mu^\mathrm{A})$ and $(\lambda^\mathrm{B}, \mu^\mathrm{B})$, respectively, and let $(\lambda (\bm{x}), \mu (\bm{x}))$ denote the spatially varying Lam\'{e}'s constants, which take either $(\lambda^\mathrm{A}, \mu^\mathrm{A})$ or $(\lambda^\mathrm{B}, \mu^\mathrm{B})$ depending on the coordinate $\bm{x}$.
The constitutive equation derived by the Hooke's law is
\begin{align}
    \sigma_{ij}^\epsilon(\bm{x}) &= C_{ijkl}^\epsilon(\bm{x}) \varepsilon_{kl}(\bm{x}) \nonumber \\
    &= \lambda(\bm{x}) \delta_{ij} \varepsilon_{kk}(\bm{x}) + 2\mu \varepsilon_{ij}(\bm{x}), \label{eq:constitutive_equation}
\end{align}
where we use the relationship between an elastic tensor $C_{ijkl}$ of an isotropic material and its Lam\'{e}'s constants $(\lambda, \mu)$ of $C_{ijkl} = \lambda \delta_{ij} \delta_{kl} + \mu (\delta_{ik} \delta_{jl} + \delta_{il} \delta_{jk})$ with the Kronecker's delta $\delta_{ij}$.
The strain tensor $\varepsilon_{ij}(\bm{x})$ is given by the strain-displacement relationship:
\begin{align}
    \varepsilon_{ij}(\bm{x}; \bm{u}^\epsilon) = \frac{1}{2} \left( \pdif{ u_{i}^\epsilon (\bm{x})}{x_{j}} + \pdif{ u_{j}^\epsilon (\bm{x})}{ x_{i}} \right), \label{eq:strain_displacement_relationship}
\end{align}
where $\bm{u}^\epsilon(\bm{x}) = (u_0^\epsilon(\bm{x}) \dots, u_{d-1}^\epsilon(\bm{x}))^\top$ is the displacement vector.
Together with Eqs.~\eqref{eq:stress_equilibrium}--\eqref{eq:strain_displacement_relationship}, we derive the governing equation in the displacement formulation, as follows:
\begin{align}
    \delta_{ij} \pdif{}{x_j} \left( \lambda(\bm{x})  \pdif{u_{k}^\epsilon}{x_k} \right) + \pdif{}{x_j} \left( \mu(\bm{x}) \left( \pdif{u_{i}^\epsilon (\bm{x})}{x_j} + \pdif{u_{j}^\epsilon (\bm{x})}{x_i} \right) \right) + f_i^\epsilon(\bm{x}) = 0. \label{eq:displacement_formulation}
\end{align}
Similar to the case of thermal conduction, the asymptotic homogenization technique enables us to estimate the effective elastic tensor $\overline{C}_{ijkl}$, as follows~\cite{chung2001asymptotic, kalamkarov2009asymptotic}:
\begin{align}
    \overline{C}_{ijkl} = \frac{1}{| Y |} \int_Y \left( \lambda(\bm{y}) \delta_{ij} \delta_{k'l'} + \mu(\bm{y}) (\delta_{ik'} \delta_{jl'} + \delta_{il'} \delta_{jk'}) \right) \left( I^{kl}_{k'l'} - \pdif{\xi^{kl}_{k'}(\bm{y})}{y_{l'}} \right) \mathrm{d} \bm{y},
    \label{eq:homo_elastic}
\end{align}
where $I^{kl}_{k'l'}:= ( \delta_{kk'}\delta_{ll'} + \delta_{lk'}\delta_{kl'} )/2$, and $\bm{\xi}^{kl}(\bm{y})=[\xi^{kl}_{0}(\bm{y}), \dots, \xi^{kl}_{d-1}(\bm{y})]^\top$ is the characteristic displacement field satisfying
\begin{align}
    \begin{cases}
        & \pdif{}{y_{i}} \left( \delta_{ij} \lambda(\bm{y}) \pdif{\xi_{k'}^{kl}(\bm{y})}{y_{k'}} + \mu(\bm{y}) \left( \pdif{\xi_{i}^{kl}(\bm{y})}{y_{j}} + \pdif{\xi_{j}^{kl}(\bm{y})}{y_{i}} \right) \right) \\
        &= \delta_{ij} \delta_{kl} \pdif{\lambda(\bm{y})}{y_{i}} + (\delta_{ik} \delta_{jl} + \delta_{il} \delta_{jk}) \pdif{\mu(\bm{y})}{y_{i}} \text{ in } Y \\
        & \bm{\xi}^{kl}(\bm{y}) ~ : ~ Y \text{-periodic},
    \end{cases}
    \label{eq:cp_elastic}
\end{align}
for $k, l \in 0, \dots, d-1$.
Owing to the symmetry that $k$ and $l$, Eq.~\eqref{eq:cp_elastic} gives the relationship that $\bm{\xi}^{kl} = \bm{\xi}^{lk}$.
Therefore, if suffices to solve Eq.~\eqref{eq:cp_elastic} for $k \leq l$.
The characteristic displacement field $\bm{\xi}^{kl}$ represents the displacement induced by a unit macroscopic strain $\varepsilon_{kl}(\bm{x}; \bm{u})$.
The asymptotic homogenization technique consists of two steps: we first solve Eq.~\eqref{eq:cp_elastic} in the RVE to obtain the characteristic displacement field $\bm{\xi}^{kl}(\bm{y})$, and then calculate the homogenized elastic tensor by Eq.~\eqref{eq:homo_elastic}.

\subsection*{Tensor train format}
The tensor-train (TT) format~\cite{oseledets2009compact, oseledets2011tensor, holtz2012manifolds, holtz2012alternating} is a representation scheme of high-dimensional arrays, or tensors, and is mainly used in computational science. 
It is particularly useful for representing high-dimensional tensors in a compressed manner.
The TT format is also known as matrix product states (MPSs) and matrix product operators (MPOs)~\cite{vidal2003efficient, schollwock2011density}, which are components of tensor networks~\cite{orus2014practical, montangero2018introduction, orus2019tensor}, in quantum physics.

The TT format of a $2m$-order tensor $\bm{A} \in \mathbb{R}^{d_{m-1} \times \dots \times d_0 \times d'_{m-1} \times \dots \times d'_0}$ is given as
\begin{align} \label{eq:tensor-train}
    &A_{p_{m-1} \dots p_0 q_{m-1} \dots q_0} \nonumber \\
    &= \sum_{\alpha_{m}=0}^{0} \sum_{\alpha_{m-1}=0}^{r_{m-1}-1} \dots \sum_{\alpha_1=0}^{r_1 - 1} \sum_{\alpha_0=0}^{0} A^{[m-1]}_{\alpha_m p_{m-1} q_{m-1} \alpha_{m-1}} \dots A^{[0]}_{\alpha_1 p_0 q_0 \alpha_0},
\end{align}
where $A_{p_{m-1} \dots, p_0 q_{m-1} \dots, q_0}$ is the component of $\bm{A}$, $\bm{A}^{[s]} \in \mathbb{R}^{r_{s+1} \times d_s \times d'_s \times r_{s}}$ is called the core of the TT format, and $r_s$ is called the TT rank, also called the bond dimension in tensor networks.
We herein set $r_m=r_0=1$.
Note that the TT format in Eq.~\eqref{eq:tensor-train} corresponds to the MPO when $d_s = d'_s$ for all $s$, and corresponds to the MPS when $d'_s=1$ for all $s$.
When $d_s=2$ for all $s$, which is a typical case in quantum physics, we particularly call the TT format the quantized tensor train (QTT).
For readability, we simply notate the TT format in Eq.~\eqref{eq:tensor-train}, as follows:
\begin{align}
    A_{p_{m-1} \dots p_0 q_{m-1} \dots q_0} &= \bm{A}^{[m-1]}_{p_{m-1} q_{m-1}} \cdots \bm{A}^{[0]}_{p_0 q_0},
\end{align}
where $\bm{A}^{[s]}_{p_s q_s}$ is the matrix whose $(\alpha_{s+1}, \alpha_{s})$-component is $A^{[s]}_{\alpha_{s+1} p_s q_s \alpha_{s}}$.
Arbitrary tensors $\bm{A} \in \mathbb{R}^{d_m \times \dots \times d_1 \times d'_m \times \dots \times d'_1}$ can be exactly represented by the TT format by the successive application of the singular value decomposition (SVD)~\cite{holtz2012manifolds}.
Also, an arbitrary matrix $\bm{M} \in \mathbb{R}^{\prod_{i=0}^{m-1} d_i \times \prod_{j=0}^{m-1} d'_j}$ with the $(i, j)$-component $M_{ij}$ can be represented in the TT format by associating this matrix with a $2m$-order tensor $\bm{A}$ by $A_{p_{m-1} \dots p_0 q_{m-1} \dots q_0} = M_{ij}$ where $i := \sum_{l=0}^{m-1} p_l \prod_{k=0}^{l-1} d_k$ and $j := \sum_{l=0}^{m-1} q_l \prod_{k=0}^{l-1} d'_k$, as described in Supplementary information~S1.
By truncating the ranks $r_s$ to a small value at each SVD, we can also obtain the compressed TT format where the truncation is locally optimal at each SVD but may not be globally optimal~\cite{paeckel2019time}.
Representing the spatial derivative and the spatially varying material properties in the TT format, we can apply TT-based numerical approaches to solve the problems~\eqref{eq:cp_thermal} and \eqref{eq:cp_elastic} defined on RVEs.
We validate our proposed method by applying it to a microstructure whose homogenized properties can be analytically estimated, which is provided in Supplementary Information~S2.

\subsection*{Effect of the volume fraction of microstructures on the computational efficiency}

We examine the effect of the volume fraction, which is an important aspect of microstructures, on computational efficiency.
We applied the asymptotic homogenization of the thermal conductivity tensor and the elastic tensor for randomly generated RVEs using the Voronoi tessellation, which mimics the polycrystalline dual-phase composite, by the procedure described in Supplementary information~S3.
We searched for the lowest value of the maximum rank of TT to estimate the homogenized material tensor within the relative error of 0.01 to those calculated by the full-rank FDM for high-resolution RVE.
Here, we used a homogenized material tensor by the full-rank FDM as a reference.
For randomly generating RVEs by the Voronoi tessellation, we generated 100 points uniformly at random in RVEs and assigned a binary 0 or 1 to each point where the probability of assigning 0, was 0.5, 0.7, or 0.9, which corresponds to the volume fraction.
Using the notation in Supplementary information~S3, these are $N_\mathrm{point}=100$, $V_\mathrm{f}=0.5,~0.7,~0.9$.
We generated 10 RVEs in both two- and three dimensions which are shown in Supplementary information~S3 and estimated their homogenized thermal conductivity and elastic tensor.
We set the thermal conductivity of each phase to $\kappa^\mathrm{A}=1$ and $\kappa^\mathrm{B}=0.5$ and set the Young's modulus of each phase to $E^\mathrm{A}=1$ and $E^\mathrm{B}=0.5$ while we set the Poisson ratio to 0.3.

\begin{figure}[t]
    \centering
    \includegraphics[width=1.0\textwidth]{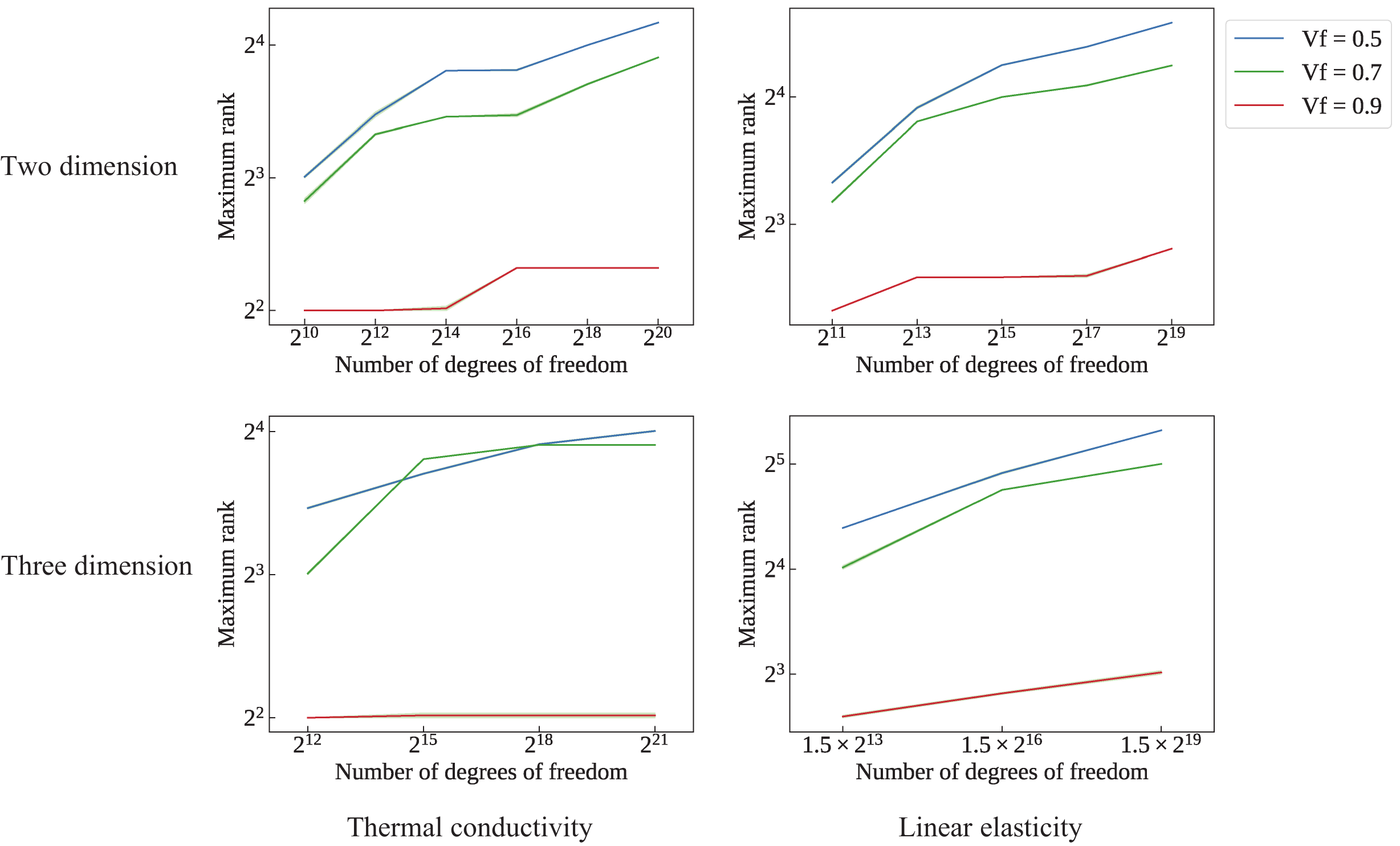}
    \caption{The lowest value of the maximum rank of TT to estimate the homogenized tensors of ten randomly generated RVEs within the error of 0.01 for various values of the number of degrees of freedom. Lines represent the medians and shaded areas show the first and third quantiles for 10 RVEs.}
    \label{fig:result_random_cell}
\end{figure}

\begin{figure}[t]
    \centering
    \includegraphics[width=1.0\textwidth]{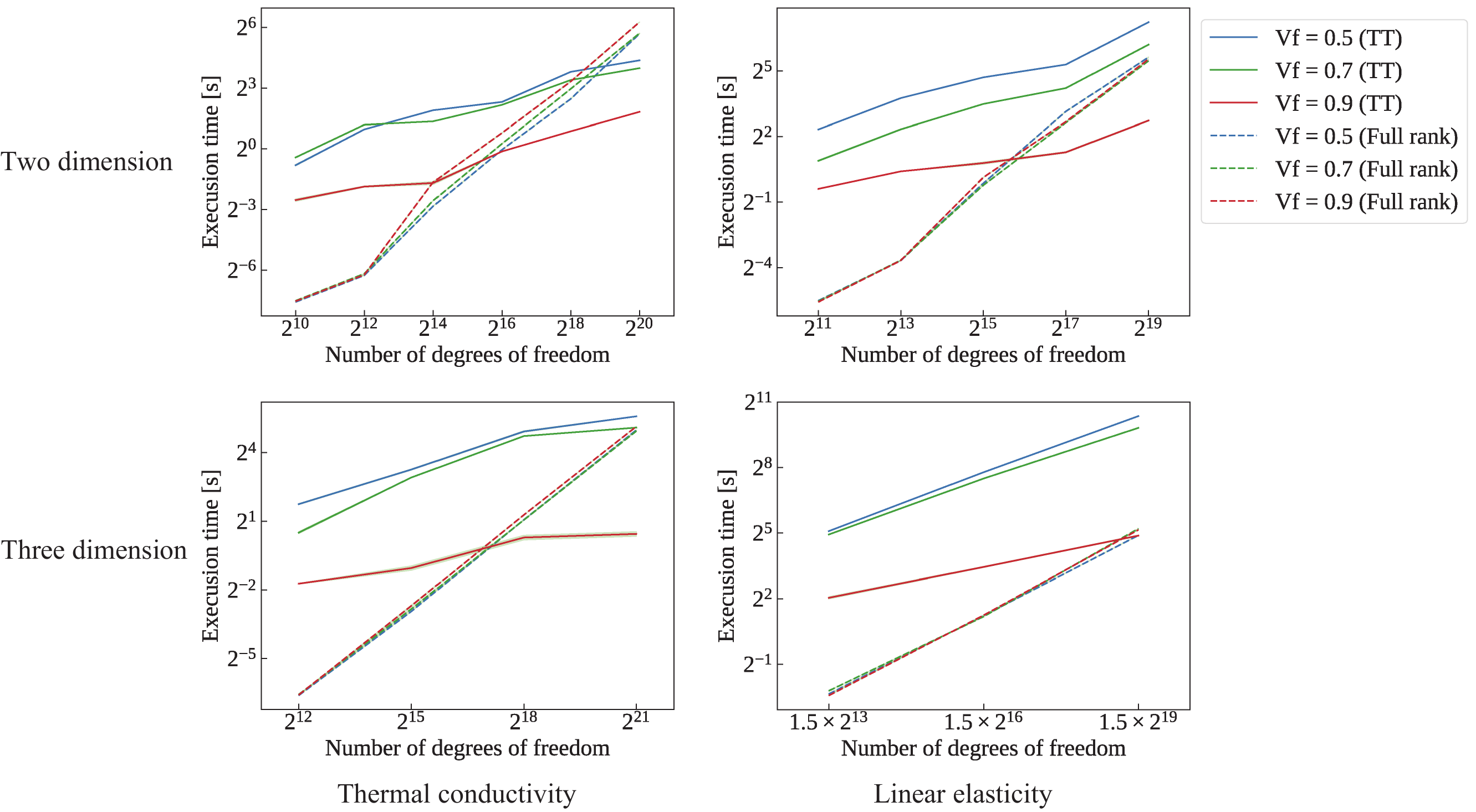}
    \caption{The computational time required to estimate the homogenized tensors of ten randomly generated RVEs using the number of ranks achieving an error lower than 0.01 for various values of the number of degrees of freedom. Lines represent the medians and shaded areas show the first and third quantiles for 10 RVEs.}
    \label{fig:result_random_cell_time}
\end{figure}

Figure~\ref{fig:result_random_cell} illustrates the lowest value of the maximum rank of TT versus the number of degrees of freedom of the asymptotic homogenization for various settings of the volume fraction of randomly generated RVEs.
This figure clearly shows that the higher the volume fraction of RVEs is, the lower the required number of TT ranks is to accurately estimate the homogenized material tensors, which implies that microstructures with a skewed composition ratio can be efficiently represented by TT with low ranks.
This trend applies to all cases of two-dimensional and three-dimensional thermal conduction and linear elasticity.
It is worth noting that the TT rank for the case of the thermal conduction problem for both two and three dimensions is significantly low and almost independent of the problem size when $V_\mathrm{f}=0.9$.
In contrast, the TT rank for the case of the linear elasticity problem for three dimensions depends on the problem size almost polynomially.
This implies that the use of the TT format does not bring significant computational benefit for estimating the elastic property of materials with heterogeneity in three dimensions.

Figure~\ref{fig:result_random_cell_time} shows the computational time required to estimate the homogenized material properties versus the number of degrees of freedom of the asymptotic homogenization for various settings of the volume fraction of randomly generated RVEs.
In all cases, the increase of the computational time of the proposed method (denoted by TT) as the increase of the degree of freedom is gradual compared to that of the conventional finite difference method (FDM), which we call full-rank FDM.
Therefore, the proposed method can bring computational benefits for large-scale problems.
However, the overhead of the proposed method is much higher than that of the full-rank FDM, especially in the case of the linear elasticity problem for three dimensions.
This overhead depends on the maximum rank of TT required for accurate calculation.
For handling the vector field involved in the linear elastic problem, we used the TT format, the rightmost core of which corresponds to each component of the vector field while other cores correspond to the spatial coordinate as we describe in Method section.
This would result in the long-range interaction of TT, which leads to large TT ranks compared to the case of thermal conduction problems.
We would like to explore other formats widely used as tensor networks~\cite{orus2014practical, montangero2018introduction, orus2019tensor, ran2020tensor} in our future works.

\section*{Discussion} \label{sec:discussion}
Our results clearly show the potential of the tensor-train (TT) format in enhancing the efficiency of computational homogenization for multiscale systems. 
By leveraging the TT format, we have demonstrated its effectiveness in estimating the homogenized properties of heterogeneous materials for complex, randomly generated microstructures.

We explored the influence of microstructural volume fractions on the efficiency of the TT-based method, using the randomly generated RVEs by Voronoi tessellation. 
By observing the TT ranks required to achieve a certain accuracy for RVEs with various volume fractions, we demonstrated that microstructures with higher volume fractions (e.g., $V_\mathrm{f}=0.9$) require fewer TT ranks compared to those with more balanced compositions (e.g., $V_\mathrm{f}=0.5$), as illustrated in Fig.~\ref{fig:result_random_cell}.
This is a significant observation as it suggests that the TT format is particularly suited for materials with pronounced heterogeneities, where certain phases dominate.
This aligns with previous studies showing that tensor networks can effectively capture localized interactions~\cite{gourianov2022quantum, kiffner2023tensor}, analogous to those in quantum many-body systems. 
As shown in Fig.~\ref{fig:result_random_cell_time}, the computational time for the proposed method increases gradually with the degree of freedom compared to that of the conventional finite difference method, which implies that the proposed method can bring computational benefit for large-scale problems.
Thus, our findings support the TT format as a powerful tool for bridging the computational gap in multiscale material analysis.

Despite these positive results, this study has certain limitations. 
Although the proposed method exhibited good scaling with respect to the increase in the number of degrees of freedom, its actual computational time can be much longer than that of the conventional FDM, especially in the case of the linear elasticity problem for three dimensions.
As we mentioned in the Result section, it would be because of the long-range interaction of TT for handling the vector field involved in the linear elastic problem, which leads to large TT ranks compared to the case of thermal conduction problems.
We would like to explore other tensor networks, including projected entangled pair states (PEPS)~\cite{verstraete2004renormalization, orus2014practical} and tree tensor networks (TTN)~\cite{shi2006classical, ke2023tree}, to effectively deal with the long-range interaction in our future works.
Furthermore, the TT-based asymptotic homogenization method was evaluated only on thermal conductivity and elasticity problems.
While these are common in material science, further validation is needed for other physics and material systems, including non-linearity. 
We are also interested in integrating the TT format with more sophisticated but computationally expensive computational homogenization techniques, such as the coupling with molecular dynamics simulation~\cite{mortazavi2013combined, terashima2024fine}, which could further enhance its efficiency and broaden its applicability in computational material science.

In conclusion, our study highlights the significant benefits of using the TT format for computational homogenization, offering an efficient and scalable method for multiscale material analysis. 
We believe that this approach opens new avenues for exploring complex material properties in a computationally efficient manner.

\section*{Methods}
In this paper, we use the index starting from zero because we often use the binary representation of numbers, and counting from zero fits the binary representation.

\subsection*{Finite difference operators}
To numerically solve the partial differential equations such as Eqs.~\eqref{eq:cp_thermal} and \eqref{eq:cp_elastic}, we have to discretize the differential operators.
Here, we use the finite difference method (FDM).
Let us consider a one-dimensional closed domain $[0, L]$ where $L$ is the length of the domain.
We discretize the domain by uniformly distributed $N + 1$ points with the interval $h := L / N$, i.e., the coordinate of the $i$-th points is $x_i = hi$ where we count $i$ from zero to $N$.
Let $u(x)$ denote the scalar value at the position $x \in [0, L]$.
The finite difference method discretizes the scalar field $u$ using its value on the $N + 1$ points; the discretized scalar field is represented by $\tilde{u}_i := u(x_i)$.
The central difference scheme approximates the first-order difference as
\begin{align}
    \left. \pdif{u}{y} \right|_{x = x_i} = \frac{\tilde{u}_{i + 1} - \tilde{u}_{i - 1}}{2h} + o(h^2) \text{ for } i = 0, \cdots, N,
\end{align}
where $\tilde{u}_{-1}$ for $i=0$ and $\tilde{u}_{N+1}$ for $i=N$ are determined from the boundary condition (BC).
In this study, we consider the periodic boundary condition, which imposes $u(0) = u(L)$, i.e., $\tilde{u}_0 = \tilde{u}_N$ and thus $\tilde{u}_{-1} = \tilde{u}_{N - 1}$ and $\tilde{u}_{N+1} = \tilde{u}_1$.
In such a scenario, we no longer need to consider $\tilde{u}_N$ as a variable and it suffices to deal with $N$ points consisting of $x_0, \dots x_{N-1}$.
Hence, we define a variable to represent the discretized scalar field as $\tilde{\bm{u}} := (\tilde{u}_0, \tilde{u}_1, \cdots, \tilde{u}_{N-1})$.
The central difference operator under the periodic boundary condition can be represented by a sparse matrix of $N \times N$, denoted by $D$, as follows:
\begin{align}
    \bm{D} := \begin{pmatrix}
        0 & 1 & 0 & \cdots & 0 & -1 \\
        -1 & 0 & 1 & \cdots & 0 & 0 \\
        0 & -1 & 0 & 1 & \cdots & 0 \\
        \vdots &  & \ddots & & \ddots & \vdots \\
        0 & \cdots & 0 & -1 & 0 & 1 \\
        1 & 0 & \cdots & 0 & -1 & 0
    \end{pmatrix}.
\end{align}
Actually, it acts on $\tilde{\bm{u}}$ as
\begin{align}
    \left(\bm{D} \tilde{\bm{u}} \right)_i = \frac{\tilde{u}_{i + 1} - \tilde{u}_{i - 1}}{2h}.
\end{align}

\begin{figure}[t]
    \centering
    \includegraphics[width=0.4\textwidth]{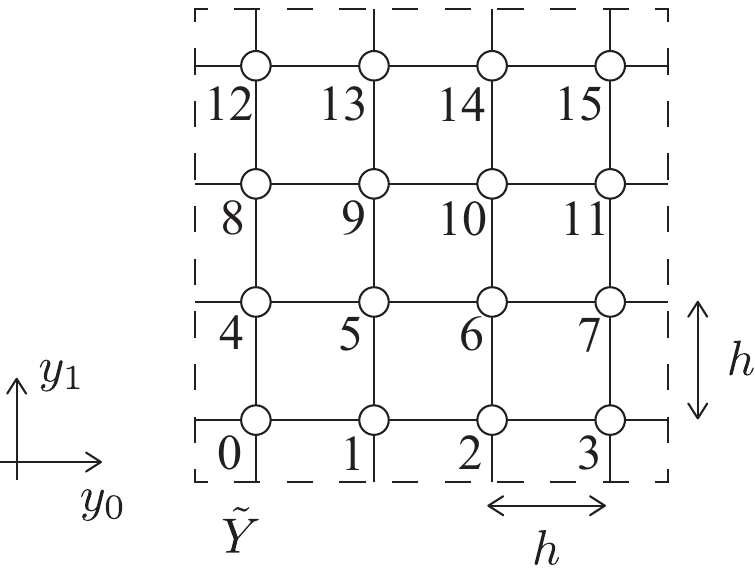}
    \caption{The lattice $\tilde{Y}$ in two dimensions ($d=2$) when $N=4$.}
    \label{fig:lattice_2d}
\end{figure}

Let the representative volume element (RVE) $Y$ be a square in two dimensions or a cube in three dimensions and be discretized by a uniform lattice, denoted by $\tilde{Y}$, of $(N + 1)^d$ nodes with the interval spacing $h:=1 / N$ where $d(=2, 3)$ is the number of spatial dimensions.
Since nodes on the edge of the lattice coincide with nodes on the edge on the opposite side due to the periodicity of the RVE $Y$, we only consider the partial lattice consisting of $N^d$ independent nodes.
Let $\{ \bm{e}_j \}_{j=0}^{d-1}$ be the orthonormal basis along with the edges of the RVE $Y$.
We number the $g_j$-th node along with the $y_j$-axis as the $g=\sum_{j=0}^{d-1} N^{j} g_j$ as shown in Fig.~\ref{fig:lattice_2d}.
Then, the first-order derivative along with the $y_j$-axis, denoted by $\bm{D}^{(y_j)}$, is represented as
\begin{align} \label{eq:central_diff_operator}
    \bm{D}^{(y_j)} = I^{\otimes (d - j - 1)} \otimes \bm{D} \otimes I^{\otimes j},
\end{align}
where $\otimes$ represents the Kronecker's product.
That is, $\bm{D}^{(y_j)}$ is the sparse matrix with the size of $N^d \times N^d$.
In the subsequent sections, we introduce the tensor-train format to efficiently represent the difference operator and a scalar field (also a vector field) discretized on the lattice $\tilde{Y}$.

\subsection*{Finite difference operators in quantized tensor-train format}
Here, we describe the finite difference operators with the periodic boundary condition in the QTT format based on the work by Kiffner and Jaksch~\cite{kiffner2023tensor}.
We first introduce the following two $2 \times 2$ matrices:
\begin{align} \label{eq:sigma_operators}
    \sigma^{01} := \begin{pmatrix}
        0 & 1 \\
        0 & 0
    \end{pmatrix}, ~
    \sigma^{10} := \begin{pmatrix}
        0 & 0 \\
        1 & 0
    \end{pmatrix}.
\end{align}
We can represent the central difference operator with the periodic boundary condition in Eq.~\eqref{eq:central_diff_operator} in the QTT format, as follows:
\begin{align}
    &D^{(y_j)}_{p_{dn-1} \dots p_0 q_{dn-1} \dots q_0} \nonumber \\
    &:= \bm{L} \left( \prod_{s=dn-1}^{(j+1)n} \bm{I}_{p_{s} q_{s}} \right) \left( \prod_{s=(j+1)n-1}^{jn} \bm{D}_{p_{s} q_{s}} \right) \left( \prod_{s=jn-1}^{0} \bm{I}_{p_{s} q_{s}} \right) \bm{R},
\end{align}
where $\bm{D}_{p_{s} q_{s}}$, $\bm{I}_{p_{s} q_{s}}$, $\bm{L}$ and $\bm{R}$ are the cores in the QTT format defined as
\begin{align}
    \bm{D}_{p_{s} q_{s}} &:=
        \begin{pmatrix}
            \delta_{p_{s} q_{s}} & \sigma^{01}_{p_{s} q_{s}} & \sigma^{10}_{p_{s} q_{s}} & 0 & 0 \\
            0 & \sigma^{10}_{p_{s} q_{s}} & 0 & 0 & 0 \\
            0 & 0 & \sigma^{01}_{p_{s} q_{s}} & 0 & 0 \\
            0 & 0 & 0 & \sigma^{10}_{p_{s} q_{s}} & 0 \\
            0 & 0 & 0 & 0 & \sigma^{01}_{p_{s} q_{s}}
        \end{pmatrix} \\
    \bm{I}_{p_{s} q_{s}} &:=
        \begin{pmatrix}
            \delta_{p_{s} q_{s}} & 0 & 0 & 0 & 0 \\
            0 & \delta_{p_{s} q_{s}} & 0 & 0 & 0 \\
            0 & 0 & \delta_{p_{s} q_{s}} & 0 & 0 \\
            0 & 0 & 0 & \delta_{p_{s} q_{s}} & 0 \\
            0 & 0 & 0 & 0 & \delta_{p_{s} q_{s}}
        \end{pmatrix} \\
    \bm{L} &:= \frac{1}{2h} \begin{pmatrix}
        1 & 0 & 0 & 1 & 1
    \end{pmatrix} \\
    \bm{R} &:= \begin{pmatrix}
        0 \\ 
        1 \\
        -1 \\
        1 \\
        -1
    \end{pmatrix}.
\end{align}
The lower-right $2 \times 2$ components of the core contribute to the periodic boundary condition.
The value $D^{(y_j)}_{p_{dn-1} \dots p_0 q_{dn-1} \dots q_0}$ corresponds to the $((p_{dn-1} \dots p_0)_2, (q_{dn-1} \dots q_0)_2)$-component of the matrix $\bm{D}^{(y_j)}$ in Eq.~\eqref{eq:central_diff_operator}.
For example, when $d=2$, $g=(p_{2n-1}, \dots, p_{n}, p_{n-1}, \dots, p_0)_2$ corresponds to the node number in Fig.~\ref{fig:lattice_2d} where $(g_0, g_1)$ with $g_0 = (p_{n-1}, \dots, p_0)_2$ and $g_1 = (p_{2n-1}, \dots, p_{n})_2$ is the coordinate of the node.
That is, this representation in the QTT format relates the node number to the binary variables $(p_{dn-1} \dots p_0)_2$ to which the binary representation of each coordinate $(p_{(j+1)n-1} \dots p_{jn})_2$ along the $y_j$-axis are concatenated.
Along with each axis, the least significant bit corresponds to the minimal scale while the most significant bit the maximum scale.

\subsection*{Spatially varying material properties in quantized tensor-train format}
To represent spatially varying material constants, it is useful to introduce a characteristic function $\chi \in L^\infty(Y)$ defined as
\begin{align} \label{eq:chi}
    \chi(\bm{y}) := \begin{cases}
        1 & \text{ for } \bm{y} \in Y^\mathrm{A} \\
        0 & \text{ for } \bm{y} \in Y^\mathrm{B},
    \end{cases}
\end{align}
where $L^\infty(Y)$ is the Lebesgue space.
The spatially varying thermal conductivity $\kappa(\bm{y})$ of the composite at the coordinate $\bm{y}$ can be expressed using the characteristic function, as follows:
\begin{align} \label{eq:kappa_chi}
    \kappa(\bm{y}) = \kappa^\mathrm{A} \chi(\bm{y}) + \kappa^\mathrm{B} (1 - \chi(\bm{y})).
\end{align}
Similarly, we represent the spatially varying Lam\'{e}'s constants of the composite, $(\lambda, \mu)$, as follows:
\begin{align} 
    &\lambda(\bm{y}) = \lambda^\mathrm{A} \chi (\bm{y}) + \lambda^\mathrm{B} (1 - \chi (\bm{y})), \label{eq:lam_chi} \\
    &\mu(\bm{y}) = \mu^\mathrm{A} \chi (\bm{y}) + \mu^\mathrm{B} (1 - \chi (\bm{y})). \label{eq:mu_chi}
\end{align}

Now, we consider to represent the material distribution $\chi(\bm{y})$ in Eq.~\eqref{eq:chi} by the quantized tensor-train (QTT) format.

Now, let $N$ be the power of two, i.e., $N=2^n$ for $n \in \mathbb{N}$ and $g = (p_{dn-1}, \dots, p_{d(n-1)}, \dots, p_{n-1}, \dots, p_0)_2$ be the binary representation of the node number $g$ depending on the QTT format of the difference operator.
We represent the characteristic function $\chi(\bm{y})$ in Eq.~\eqref{eq:chi} on nodes of the lattice $\tilde{Y}$ by the QTT format, as follows:
\begin{align}
    \chi( \bm{y}_g ) &=: \tilde{\chi}_{p_{dn-1} \dots p_0} \nonumber \\
    &= \tilde{\bm{\chi}}^{[dn-1]}_{p_{dn-1}} \cdots \tilde{\bm{\chi}}^{[0]}_{p_0},
\end{align}
where $\bm{y}_g$ is the coordinate of the $g$-th node, $\tilde{\chi}$ is the $dn$-order tensor whose components correspond to the characteristic function value on the lattice $\tilde{Y}$, and $\tilde{\bm{\chi}}^{[s]}$ is the core of the QTT format.
Here, we set $d'_s = 1$ for $s = 1, \dots dn$, which makes the index $q_s$ take only a single value in the TT format in Eq.~\eqref{eq:tensor-train}.
Thus, we omit the index $q_s$.

\subsection*{Material homogenization by tensor-train format}
Here, we discretize the cell problems in Eqs.~\eqref{eq:cp_thermal} and \eqref{eq:cp_elastic} on the lattice $\tilde{Y}$.
The thermal conductivity  in the RVE is given in the QTT format as
\begin{align}
    \tilde{\kappa}_{p_{dn-1} \dots p_0} &= (\kappa^\mathrm{A} - \kappa^\mathrm{B}) \tilde{\chi}_{p_{dn-1} \dots p_0} + \kappa^\mathrm{B} \bm{1}_{p_{dn-1}} \cdots \bm{1}_{p_{0}} \nonumber \\
    &= \tilde{\bm{\kappa}}^{[dn-1]}_{p_{dn-1}} \cdots \tilde{\bm{\kappa}}^{[0]}_{p_{0}},
\end{align}
where $\tilde{\kappa}_{p_{dn-1}, \dots, p_0} = \kappa( \bm{y}_g )$ with $g = (p_{dn-1}, \dots, p_0)_2$, $\bm{1}_{p_{s}} = 1$ for arbitrary $s$, and we used the formula for addition in the QTT format in Supplementary information~S4.
Similarly, we obtain the QTT format of the Lam\'{e}'s constants, as follows:
\begin{align}
    \tilde{\lambda}_{p_{dn-1} \dots p_0} &= (\lambda^\mathrm{A} - \lambda^\mathrm{B}) \tilde{\chi}_{p_{dn-1} \dots p_0} + \lambda^\mathrm{B} \bm{1}_{p_{dn-1}} \cdots \bm{1}_{p_{0}} \nonumber \\
    &= \tilde{\bm{\lambda}}^{[dn-1]}_{p_{dn-1}} \cdots \tilde{\bm{\lambda}}^{[0]}_{p_{0}}, \\
    \tilde{\mu}_{p_{dn-1} \dots p_0} &= (\mu^\mathrm{A} - \mu^\mathrm{B}) \tilde{\chi}_{p_{dn-1} \dots p_0} + \mu^\mathrm{B} \bm{1}_{p_{dn-1}} \cdots \bm{1}_{p_{0}} \nonumber \\
    &= \tilde{\bm{\mu}}^{[dn-1]}_{p_{dn-1}} \cdots \tilde{\bm{\mu}}^{[0]}_{p_{0}},
\end{align}
where $\tilde{\lambda}_{p_{dn-1}, \dots, p_0} = \lambda( \bm{y}_g )$ and $\tilde{\mu}_{p_{dn-1}, \dots, p_0} = \mu( \bm{y}_g )$.

\subsection*{Cell problem for thermal conductivity tensor}
We herein discretize the cell problem in Eq.~\eqref{eq:cp_thermal} onto the lattice $\tilde{Y}$.
First, we approximate the derivative in the left-hand side of Eq.~\eqref{eq:cp_thermal} by the central difference scheme, as follows:
\begin{align} \label{eq:cp_thermal_left_approx}
    &\left. \pdif{}{y_i} \left( \kappa(\bm{y}) \pdif{\phi^{j} (\bm{y})}{y_i} \right) \right|_{\bm{y}=\bm{y}_{g}} \nonumber \\
    &= \sum_{i=1}^d D^{(y_i)}_{p_{dn-1} \dots p_0 q'_{dn-1} \dots q'_0} \mathrm{diag} \left( \tilde{\bm{\kappa}} \right)_{p'_{dn-1} \dots p'_0 q'_{dn-1} \dots q'_0} D^{(y_i)}_{p'_{dn-1} \dots p'_0 q_{dn-1} \dots q_0} \tilde{\phi}^j_{q_{dn-1} \dots q_{0}} + o(h^2),
\end{align}
where $\mathrm{diag}(\tilde{\bm{\kappa}})$ is diagonalized QTT defined as
\begin{align}
    \mathrm{diag} \left( \tilde{\bm{\kappa}} \right)_{p'_{dn-1} \dots p'_0 q'_{dn-1} \dots q'_0} = 
    \begin{cases}
        \tilde{\kappa}_{p'_{dn-1} \dots p'_0} & \text{if } \forall s \in \{0, \dots, dn-1 \}, ~ p'_s = q'_s \\
        0 & \text{otherwise}.    
    \end{cases}
\end{align}
We also use the central difference scheme to approximate the derivative in the right-hand side of Eq.~\eqref{eq:cp_thermal}, as follows:
\begin{align}
    &\left. \pdif{\kappa(\bm{y})}{y_j} \right|_{\bm{y}=\bm{y}_{g}} = D^{(y_j)}_{p_{dn-1} \dots p_0 q_{dn-1} \dots q_0} \tilde{\kappa}_{q_{dn-1} \dots q_0} + o(h^2).
\end{align}
Eventually, we obtain the cell problem in the QTT format, as follows:
\begin{align} \label{eq:cp_thermal_tt}
    &\sum_{i=1}^d D^{(y_i)}_{p_{dn-1} \dots p_0 q'_{dn-1} \dots q'_0} \mathrm{diag} \left( \tilde{\bm{\kappa}} \right)_{p'_{dn-1} \dots p'_0 q'_{dn-1} \dots q'_0} D^{(y_i)}_{p'_{dn-1} \dots p'_0 q_{dn-1} \dots q_0} \tilde{\phi}^j_{q_{dn-1} \dots q_{0}} \nonumber \\
    &= D^{(y_j)}_{p_{dn-1} \dots p_0 q_{dn-1} \dots q_0} \tilde{\kappa}_{q_{dn-1} \dots q_0},
\end{align}
where $\tilde{\phi}^j_{q_{dn-1} \dots q_{0}}$ is the unknown variable.
Now, since we have both the operator and the right-hand side in the QTT format, we can solve Eq.~\eqref{eq:cp_thermal_tt} by linear system solvers in the TT format.
Specifically, we employ the modified alternating least square (MALS) method~\cite{holtz2012alternating}, which sequentially optimizes the cores of the TT format of variables with dynamical rank adaptation by solving linear equations for each core.
We now evaluate the homogenized thermal conductivity tensor in Eq.~\eqref{eq:homo_thermal_tt} using the QTT format, as follows:

\begin{align}
    \overline{\kappa}_{ij} &= \frac{1}{|Y|} \int_Y \kappa(\bm{y}) \left( \delta_{ij} - \pdif{\phi^{j} (\bm{y}) }{y_{i}}  \right) \mathrm{d} \bm{y} \nonumber \\
    &\approx  h^d \left( \bm{1}_{p_{dn-1}} \cdots \bm{1}_{p_{0}} \tilde{\kappa}_{p_{dn-1} \dots p_0} \delta_{ij} - \tilde{\kappa}_{p_{dn-1} \dots p_0} D^{(y_i)}_{p_{dn-1} \dots p_0 q_{dn-1} \dots q_0} \tilde{\phi}^{j}_{q_{dn-1} \dots q_0} \right),
    \label{eq:homo_thermal_tt}
\end{align}
where $\bm{1}_{p_s} = 1$ for arbitrary $s$.

\subsection*{Cell problem for linear elastic material}
Here, we discretize the cell problem in Eq.~\eqref{eq:cp_elastic} onto the lattice $\tilde{Y}$.
In the following, we discuss the two-dimensional problems, but we can easily apply the following discussion for three-dimensional problems.
First, we expand the summation in Eq.~\eqref{eq:cp_elastic}, as follows:
\begin{align} \label{eq:cp_elastic_matrix}
    \begin{pmatrix}
        K_{00} (\bm{y}) & K_{01} (\bm{y}) \\
        K_{10} (\bm{y}) & K_{11} (\bm{y})
    \end{pmatrix} \begin{pmatrix}
        \xi^{kl}_{0} (\bm{y}) \\
        \xi^{kl}_{1} (\bm{y})
    \end{pmatrix} = \begin{pmatrix}
        F_{0}^{kl} (\bm{y}) \\
        F_{1}^{kl} (\bm{y})
    \end{pmatrix},
\end{align}
where
\begin{align}
    & K_{00} (\bm{y}) := \pdif{}{y_{0}} (\lambda(\bm{y}) + 2 \mu(\bm{y})) \pdif{}{y_{0}} + \pdif{}{y_{1}} \mu(\bm{y}) \pdif{}{y_{1}} \\
    & K_{01} (\bm{y}) := \pdif{}{y_{0}} \lambda(\bm{y}) \pdif{}{y_{1}} + \pdif{}{y_{1}} \mu(\bm{y}) \pdif{}{y_{0}} \\
    & K_{10} (\bm{y}) := \pdif{}{y_{1}} \lambda(\bm{y}) \pdif{}{y_{0}} + \pdif{}{y_{0}} \mu(\bm{y}) \pdif{}{y_{1}} \\
    & K_{11} (\bm{y}) := \pdif{}{y_{1}} (\lambda(\bm{y}) + 2 \mu(\bm{y})) \pdif{}{y_{1}} + \pdif{}{y_{0}} \mu(\bm{y}) \pdif{}{y_{0}} \\
    & F_{0}^{kl} (\bm{y}) := \delta_{kl} \pdif{\lambda(\bm{y})}{y_{0}} + \delta_{0l} \pdif{\mu(\bm{y})}{y_{k}} + \delta_{0k} \pdif{\mu(\bm{y})}{y_{l}} \\
    & F_{1}^{kl} (\bm{y}) := \delta_{kl} \pdif{\lambda(\bm{y})}{y_{1}} + \delta_{1l} \pdif{\mu(\bm{y})}{y_{k}} + \delta_{1k} \pdif{\mu(\bm{y})}{y_{l}}.
\end{align}
Then, we introduce the TT format for the vector $(\xi_0^{kl}(\bm{y}_g), \xi_1^{kl} (\bm{y}_g))^\top$ written as
\begin{align} \label{eq:xi_tt}
    \tilde{\xi}_{p_{dn-1} \dots p_0 j} = \tilde{\bm{\xi}}^{[dn]}_{p_{dn-1}} \dots \tilde{\bm{\xi}}^{[1]}_{p_{0}} \tilde{\bm{\xi}}^{[0]}_{j},
\end{align}
where $\tilde{\bm{\xi}}^{[s+1]}_{p_{s}}$ and $\tilde{\bm{\xi}}^{[0]}_{j}$ are the cores of the TT where $\tilde{\bm{\xi}}^{[0]}_{j}$ is the core for representing each component of the vector $(\xi_0^{kl}(\bm{y}), \xi_1^{kl} (\bm{y}))^\top$.
We call $\tilde{\xi}_{p_{dn-1} \dots p_0 j}$ the TT format because the TT format $\tilde{\xi}_{p_{dn-1} \dots p_0 j}$ is no longer a QTT for three-dimensional problems ($d=3$), where $j=0, 1, 2$, but we do not have to distinguish TT and QTT in the following discussion.
Now, we discretize the matrix of the left-hand side in Eq.~\eqref{eq:cp_elastic_matrix}.
Using the central difference operator, we obtain the QTT formats of each component of the matrix, as follows:
\begin{align}
    &(\tilde{K}_{00})_{p_{dn-1} \dots p_0 q_{dn-1} \dots q_0} \nonumber \\
    &= D^{(y_0)}_{p_{dn-1} \dots p_0 q'_{dn-1} \dots q'_0} \mathrm{diag} \left( \tilde{\bm{\lambda}} + 2 \tilde{\bm{\mu}} \right)_{p'_{dn-1} \dots p'_0 q'_{dn-1} \dots q'_0} D^{(y_0)}_{p'_{dn-1} \dots p'_0 q_{dn-1} \dots q_0} \nonumber \\
    &\quad + D^{(y_1)}_{p_{dn-1} \dots p_0 q'_{dn-1} \dots q'_0} \mathrm{diag} \left( \tilde{\bm{\mu}} \right)_{p'_{dn-1} \dots p'_0 q'_{dn-1} \dots q'_0} D^{(y_1)}_{p'_{dn-1} \dots p'_0 q_{dn-1} \dots q_0} \\
    &(\tilde{K}_{01})_{p_{dn-1} \dots p_0 q_{dn-1} \dots q_0} \nonumber \\
    &= D^{(y_0)}_{p_{dn-1} \dots p_0 q'_{dn-1} \dots q'_0} \mathrm{diag} \left( \tilde{\bm{\lambda}} \right)_{p'_{dn-1} \dots p'_0 q'_{dn-1} \dots q'_0} D^{(y_1)}_{p'_{dn-1} \dots p'_0 q_{dn-1} \dots q_0} \nonumber \\
    &\quad + D^{(y_1)}_{p_{dn-1} \dots p_0 q'_{dn-1} \dots q'_0} \mathrm{diag} \left( \tilde{\bm{\mu}} \right)_{p'_{dn-1} \dots p'_0 q'_{dn-1} \dots q'_0} D^{(y_0)}_{p'_{dn-1} \dots p'_0 q_{dn-1} \dots q_0} \\
    &(\tilde{K}_{10})_{p_{dn-1} \dots p_0 q_{dn-1} \dots q_0} \nonumber \\
    &= D^{(y_1)}_{p_{dn-1} \dots p_0 q'_{dn-1} \dots q'_0} \mathrm{diag} \left( \tilde{\bm{\lambda}} \right)_{p'_{dn-1} \dots p'_0 q'_{dn-1} \dots q'_0} D^{(y_0)}_{p'_{dn-1} \dots p'_0 q_{dn-1} \dots q_0} \nonumber \\
    &\quad + D^{(y_0)}_{p_{dn-1} \dots p_0 q'_{dn-1} \dots q'_0} \mathrm{diag} \left( \tilde{\bm{\mu}} \right)_{p'_{dn-1} \dots p'_0 q'_{dn-1} \dots q'_0} D^{(y_1)}_{p'_{dn-1} \dots p'_0 q_{dn-1} \dots q_0} \\
    &(\tilde{K}_{11})_{p_{dn-1} \dots p_0 q_{dn-1} \dots q_0} \nonumber \\
    &= D^{(y_1)}_{p_{dn-1} \dots p_0 q'_{dn-1} \dots q'_0} \mathrm{diag} \left( \tilde{\bm{\lambda}} + 2 \tilde{\bm{\mu}} \right)_{p'_{dn-1} \dots p'_0 q'_{dn-1} \dots q'_0} D^{(y_1)}_{p'_{dn-1} \dots p'_0 q_{dn-1} \dots q_0} \nonumber \\
    &\quad + D^{(y_0)}_{p_{dn-1} \dots p_0 q'_{dn-1} \dots q'_0} \mathrm{diag} \left( \tilde{\bm{\mu}} \right)_{p'_{dn-1} \dots p'_0 q'_{dn-1} \dots q'_0} D^{(y_0)}_{p'_{dn-1} \dots p'_0 q_{dn-1} \dots q_0} \\
    &(\tilde{F}_0)^{kl}_{p_{dn-1} \dots p_0} \nonumber \\
    &= \delta_{kl} D^{(y_0)}_{p_{dn-1} \dots p_0 q_{dn-1} \dots q_0} \tilde{\lambda}_{q_{dn-1} \dots q_0} \nonumber \\
    &\quad + \delta_{0l} D^{(y_k)}_{p_{dn-1} \dots p_0 q_{dn-1} \dots q_0} \tilde{\mu}_{q_{dn-1} \dots q_0} + \delta_{0k} D^{(y_l)}_{p_{dn-1} \dots p_0 q_{dn-1} \dots q_0} \tilde{\mu}_{q_{dn-1} \dots q_0} \\
    &(\tilde{F}_1)^{kl}_{p_{dn-1} \dots p_0} \nonumber \\
    &= \delta_{kl} D^{(y_1)}_{p_{dn-1} \dots p_0 q_{dn-1} \dots q_0} \tilde{\lambda}_{q_{dn-1} \dots q_0} \nonumber \\
    &\quad + \delta_{1l} D^{(y_k)}_{p_{dn-1} \dots p_0 q_{dn-1} \dots q_0} \tilde{\mu}_{q_{dn-1} \dots q_0} + \delta_{1k} D^{(y_l)}_{p_{dn-1} \dots p_0 q_{dn-1} \dots q_0} \tilde{\mu}_{q_{dn-1} \dots q_0},
\end{align}
where $(\tilde{K}_{jj'})_{p_{dn-1} \dots p_0 q_{dn-1} \dots q_0} = K_{jj'}(\bm{y}_g)$ and $(\tilde{F}_j)^{kl}_{p_{dn-1} \dots p_0} = F_{j}^{kl} (\bm{y}_g)$.
With the operators $\sigma^{01}$ and $\sigma^{10}$ in Eq.~\eqref{eq:sigma_operators} and operators $\sigma^{00}$ and $\sigma^{11}$ defined as 
\begin{align}
    \sigma^{00} := \begin{pmatrix}
        1 & 0 \\
        0 & 0
    \end{pmatrix}, ~ \sigma^{11} := \begin{pmatrix}
        0 & 0 \\
        0 & 1
    \end{pmatrix},
\end{align}
these components $\{ (\tilde{K}_{jj'})_{p_{dn-1} \dots p_0 q_{dn-1} \dots q_0} \}_{j,j'=0}^{1}$ and $\{ (\tilde{F}_{j})_{p_{dn-1} \dots p_0} \}_{j=0}^{1}$ are concatenated into a QTT as
\begin{align}
    \tilde{K}_{p_{dn-1} \dots p_0 j q_{dn-1} \dots q_0 j'} &= (\tilde{K}_{00})_{p_{dn-1} \dots p_0 q_{dn-1} \dots q_0} \sigma^{00}_{jj'} + (\tilde{K}_{01})_{p_{dn-1} \dots p_0 q_{dn-1} \dots q_0} \sigma^{01}_{jj'} \nonumber \\
    &\quad + (\tilde{K}_{10})_{p_{dn-1} \dots p_0 q_{dn-1} \dots q_0} \sigma^{10}_{jj'} + (\tilde{K}_{11})_{p_{dn-1} \dots p_0 q_{dn-1} \dots q_0} \sigma^{11}_{jj'} \nonumber \\
    &= \tilde{K}^{[dn]}_{p_{dn-1} q_{dn-1}} \cdots \tilde{K}^{[1]}_{p_{0} q_{0}} \begin{pmatrix}
        \sigma^{00}_{jj'} & \sigma^{01}_{jj'} & \sigma^{10}_{jj'} & \sigma^{11}_{jj'}
    \end{pmatrix}^\top \\
    \tilde{F}^{kl}_{p_{dn-1} \dots p_0 j} &= (\tilde{F}_0)^{kl}_{p_{dn-1} \dots p_0} (\bm{e}_0)_{j} + (\tilde{F}_1)^{kl}_{p_{dn-1} \dots p_0} (\bm{e}_1)_{j} \nonumber \\
    &= \tilde{F}^{kl [dn]}_{p_{dn-1}} \cdots \tilde{F}^{kl [1]}_{p_{0}} \begin{pmatrix}
        (\bm{e}_0)_{j} & (\bm{e}_1)_{j}
    \end{pmatrix}^\top,
\end{align}
where we use the addition formula in the TT format in Supplementary information~S4.
We eventually obtain the cell problem in the TT format, as follows:
\begin{align} \label{eq:cp_elastic_tt}
    &\tilde{K}_{p_{dn-1} \dots p_0 j q_{dn-1} \dots q_0 j'} \tilde{\xi}^{kl}_{q_{dn-1} \dots q_0 j'} = \tilde{F}^{kl}_{p_{dn-1} \dots p_0 j},
\end{align}
where $\tilde{\xi}^{kl}_{q_{dn-1} \dots q_{0} j'}$ is the unknown variable and is obtained by linear system solvers in the TT format.
We employ the MALS method~\cite{holtz2012alternating} to solve Eq.~\eqref{eq:cp_elastic_tt} in the same manner as Eq.~\eqref{eq:cp_thermal_tt}.

We finally approximate the homogenized tensor in Eq.~\eqref{eq:homo_elastic} by TT formats, as follows:
\begin{align}
    \overline{C}_{ijkl} &= \frac{1}{| Y |} \int_Y \left( \lambda(\bm{y}) \delta_{ij} \delta_{k'l'} + \mu(\bm{y}) (\delta_{ik'} \delta_{jl'} + \delta_{il'} \delta_{jk'}) \right) \left( I^{kl}_{k'l'} - \pdif{\xi^{kl}_{k'}(\bm{y})}{y_{l'}} \right) \mathrm{d} \bm{y} \nonumber \\
    &\approx h^d \bm{1}_{p_{dn-1}} \cdots \bm{1}_{p_{0}} \left( \tilde{\lambda}_{p_{dn-1} \dots p_0} \delta_{ij} \delta_{kl} + \tilde{\mu}_{p_{dn-1} \dots p_0} (\delta_{ik} \delta_{jl} + \delta_{il} \delta_{jk}) \right) \nonumber \\
    &\quad - h^d \left( \tilde{\lambda}_{p_{dn-1} \dots p_0} \delta_{ij} \delta_{k'l'} + \tilde{\mu}_{p_{dn-1} \dots p_0} (\delta_{ik'} \delta_{jl'} + \delta_{il'} \delta_{jk'}) \right) \nonumber \\
    & \hspace{60mm} D^{(y_{l'})}_{p_{dn-1} \dots p_0 q_{dn-1} \dots q_0} \tilde{\xi}^{kl}_{q_{dn-1} \dots q_0 k'}.
    \label{eq:homo_elastic_tt}
\end{align}

\subsection*{Procedure for homogenization analysis}
We now describe the procedure for material homogenization analysis via the tensor-train format.

\begin{enumerate}
    \item Encoding a pixel or voxel data $\chi(\bm{y})$ representing the material distribution in a RVE into the QTT format by using the scheme in Supplementary information~S1.
    Here, we truncate the ranks of the QTT by a maximum value of $\overline{r}$.
    \item Construct the operator and the right-hand side of a cell problem in Eq.~\eqref{eq:cp_thermal_tt} or \eqref{eq:cp_elastic_tt} in the QTT format.
    Here, we truncate the ranks of the QTT of operators and the right-hand side by setting a threshold of the truncation error $\varepsilon$ described in Supplementary information~S5.
    \item Solve the cell problem in Eq.~\eqref{eq:cp_thermal_tt} or \eqref{eq:cp_elastic_tt} by a linear system solver in TT format.
    Specifically, we use the MALS method~\cite{holtz2012alternating}.
    We truncate the ranks of the QTT by a maximum value of $\overline{r}$.
    \item Evaluate the homogenized material property by Eq.~\eqref{eq:homo_thermal_tt} or \eqref{eq:homo_elastic_tt}.
\end{enumerate}
All of these above procedures can be computed in linear with the number of cores $dn$ of TT and polynomial with respect to the maximum rank $\overline{r}$.
Specifically, since the time complexity of the MALS is $\mathcal{O}(\max(r_x, r_b)^3 r_A^2 n)$ for a system of linear equations $Ax=b$ where $A$, $x$ and $b$ are given by the TT format with the bond dimensions of $r_A$, $r_x$ and $r_b$, respectively~\cite{holtz2012alternating}, the time complexity for solving Eqs.~\eqref{eq:cp_thermal_tt} and \eqref{eq:cp_elastic_tt} is $\mathcal{O}(\overline{r}^5 dn)$.
On the other hand, the complexity of conventional full linear solvers scale as $\mathcal{O}(2^{dn})$ because the number of grid nodes is $2^{dn}$.
Therefore, if we can obtain a good approximation of the homogenized macroscopic properties even with the maximum ranks $\overline{r}$ in polynomial with respect to $d$ and $n$, the use of TT formats can be advantageous.

\bibliography{ref}

\end{document}


\title{Supplementary information: Efficient computational homogenization via tensor train format}

\author{Yuki Sato$^{1}$, Yuto Lewis Terashima$^1$, Ruho Kondo$^1$\\ 
$^1$Toyota Central R\&D Labs., Inc., 1-4-14, Koraku, Bunkyo-ku, Tokyo, 112-0004, Japan}
\date{}

\maketitle

\makeatletter
\renewcommand{\thesection}{S\arabic{section}}
\renewcommand{\theequation}{S\arabic{equation}}
\renewcommand{\thefigure}{S\arabic{figure}}
\renewcommand{\bibnumfmt}[1]{[S#1]}
\renewcommand{\citenumfont}[1]{S#1}

\section{Encoding a matrix into a tensor-train format} \label{sec:tt_encoding}
Let $\bm{M} \in \mathbb{R}^{\prod_{i=0}^{m-1} d_i \times \prod_{j=0}^{m-1} d'_j}$ be an arbitrary matrix with the $(i, j)$-component $M_{ij}$.
We can represent this matrix as a $2m$-order tensor $\bm{A}$ whose components are related with $M_{ij}$ as $A_{p_{m-1} \dots p_0 q_{m-1} \dots q_0} = M_{ij}$ where $i := \sum_{l=0}^{m-1} p_l \prod_{k=0}^{l-1} d_k$ and $j := \sum_{l=0}^{m-1} q_l \prod_{k=0}^{l-1} d'_k$.
Let us regard the tensor $A_{p_{m-1} \dots p_0 q_{m-1} \dots q_0}$ as the $d_{m-1} d'_{m-1} \times \prod_{i=0}^{m-2} d_i d'_i$ matrix whose components are denoted by $A_{(p_{m-1}, q_{m-1}) (q_{m-2}, q_{m-2}, \dots, p_0, q_0)}$ and apply the singular value decomposition, as follows:
\begin{align}
    &A_{(p_{m-1}, q_{m-1}) (q_{m-2}, q_{m-2}, \dots, p_0, q_0)} \nonumber \\
    &= \sum_{\alpha_{m-1}=0}^{r_{m-1}-1} U_{(p_{m-1}, q_{m-1}) \alpha_{m-1}}^{[m-1]} S_{\alpha_{m-1}}^{[m-1]} V_{(q_{m-2}, q_{m-2}, \dots, p_0, q_0) \alpha_{m-1}}^{[m-1]},
\end{align}
where $r_{m-1} := \min (d_{m-1} d'_{m-1}, \prod_{i=0}^{m-2} d_i d'_i)$, $U_{(p_{m-1}, q_{m-1}) \alpha_{m-1}}^{[m-1]}$ and $V_{(q_{m-2}, q_{m-2}, \dots, p_0, q_0) \alpha_{m-1}}^{[m-1]}$ are the unitaries and $S_{\alpha_{m-1}}^{[m-1]}$ is the singular value.
Similarly, we then regard the tensor $S_{\alpha_{m-1}}^{[m-1]} V_{(q_{m-2}, q_{m-2}, \dots, p_0, q_0) \alpha_{m-1}}^{[m-1]}$ as the $r_{m-1} d_{m-2} d'_{m-2} \times \prod_{i=0}^{m-3} d_i d'_i$ matrix whose components are denoted by $A_{(p_{m-2}, q_{m-2}, \alpha_{m-1}) (q_{m-3}, q_{m-3}, \dots, p_0, q_0)}$ and apply the singular value decomposition, as follows:
\begin{align}
    &A_{(p_{m-2}, q_{m-2}, \alpha_{m-1}) (q_{m-3}, q_{m-3}, \dots, p_0, q_0)} \nonumber \\
    &= \sum_{\alpha_{m-2}=0}^{r_{m-2}-1} U_{(\alpha_{m-1}, p_{m-2}, q_{m-2}) \alpha_{m-2}}^{[m-2]} S_{\alpha_{m-2}}^{[m-2]} V_{(q_{m-3}, q_{m-3}, \dots, p_0, q_0) \alpha_{m-2}}^{[m-2]},
\end{align}
where $r_{m-2} := \min (r_{m-1} d_{m-2} d'_{m-2}, \prod_{i=0}^{m-3} d_i d'_i)$, $U_{(\alpha_{m-1}, p_{m-2}, q_{m-2}) \alpha_{m-2}}^{[m-2]}$ and $V_{(q_{m-3}, q_{m-3}, \dots, p_0, q_0) \alpha_{m-2}}^{[m-2]}$ are the unitaries and $S_{\alpha_{m-2}}^{[m-2]}$ is the singular value.
The successive application of the singular value decomposition eventually yields
\begin{align}
    &A_{p_{m-1} \dots p_0 q_{m-1} \dots q_0} \nonumber \\
    &= \sum_{\alpha_{m-1}=0}^{r_{m-1}-1} \dots \sum_{\alpha_1=0}^{r_1 - 1}  U_{(p_{m-1}, q_{m-1}) \alpha_{m-1}}^{[m-1]} U_{(\alpha_{m-1}, p_{m-2}, q_{m-2}) \alpha_{m-2}}^{[m-2]} \dots U_{(\alpha_2, p_1, q_1) \alpha_1}^{[1]} S_{\alpha_1}^{[1]} V_{(p_0, q_0) \alpha_1}^{[1]} \nonumber \\
    &= \sum_{\alpha_{m}=0}^{0} \sum_{\alpha_{m-1}=0}^{r_{m-1}-1} \dots \sum_{\alpha_1=0}^{r_1 - 1} \sum_{\alpha_0=0}^{0} A^{[m-1]}_{\alpha_m p_{m-1} q_{m-1} \alpha_{m-1}} \dots A^{[1]}_{\alpha_2 p_1 q_1 \alpha_1} A^{[0]}_{\alpha_1 p_0 q_0 \alpha_0},
\end{align}
where we set $r_s = \min \left(r_{s+1} d_{s} d'_{s}, \prod_{i=0}^{s-1} d_i d'_i \right)$ and
\begin{align}
    \begin{cases}
        A^{[m-1]}_{0 p_{m-1} q_{m-1} \alpha_{m-1}} = U_{(p_{m-1}, q_{m-1}) \alpha_{m-1}}^{[m-1]} & \\
        A^{[s]}_{\alpha_{s+1} p_s q_s \alpha_{s}} = U_{(\alpha_{s+1}, p_s, q_s) \alpha_{s-1}}^{[s]} & \text{ for } 1 \leq s \leq m-2 \\
        A^{[0]}_{\alpha_1 p_0 q_0 0} = S_{\alpha_1}^{[1]} V_{(p_0, q_0) \alpha_1}^{[1]}. &
    \end{cases}
\end{align}
This is exactly the tensor-train format.
By truncating the rank $r_s$ by the certain value $\overline{r}$, we can obtain the TT format approximating the tensor $\bm{A}$.

\section{Comparison of the proposed method with the analytical solution} \label{sec:result_analytical}
\begin{figure}[t]
    \centering
    \includegraphics[width=0.7\textwidth]{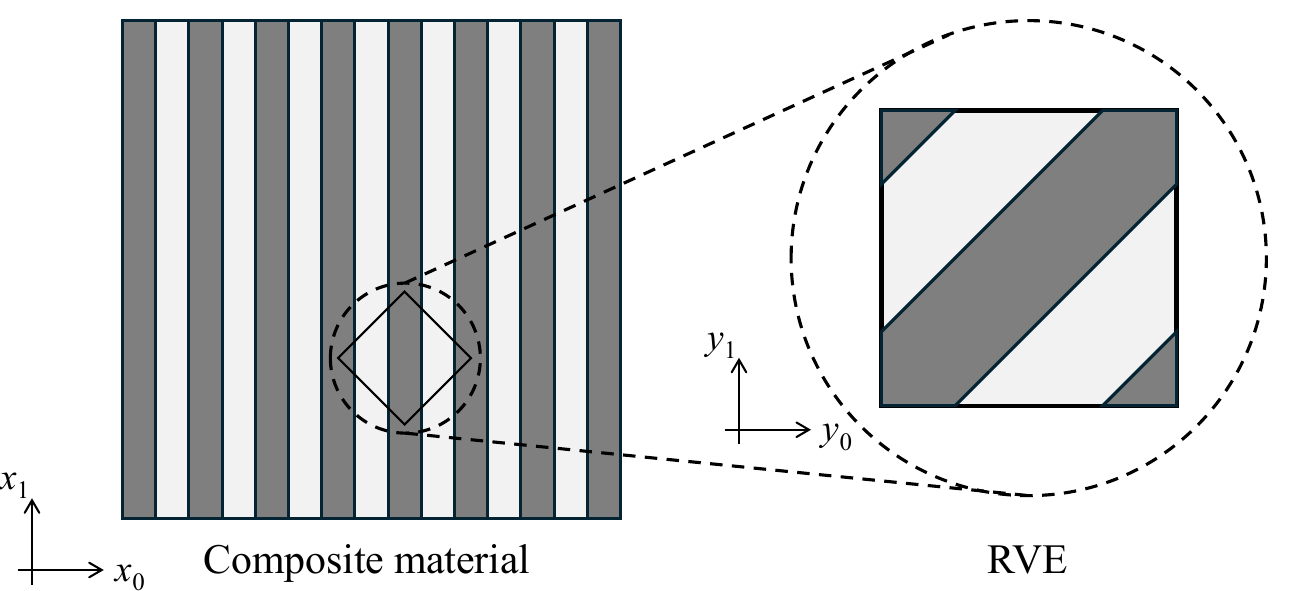}
    \caption{A schematic image of representative volume element (RVE) having the analytical homogenized thermal conductivity.}
    \label{fig:RVE_analytic_model}
\end{figure}

Here, we provide a result applying our proposed method to a microstructure whose homogenized properties can be analytically estimated.
Figure~\ref{fig:RVE_analytic_model} illustrates the representative volume element (RVE) considered here.
Let the dark gray and white regions have the thermal conductivity of $\kappa^\mathrm{A}$ and $\kappa^\mathrm{B}$, respectively.
Since this composite material has the layered property, the effective (homogenized) thermal conductivity parallel to each phase (i.e., the vertical direction in the left figure) is given as 
\begin{align}
    \kappa^\parallel := \frac{1}{2} \kappa^\mathrm{A} + \frac{1}{2} \kappa^\mathrm{B},
\end{align}
where the factor $1/2$ corresponds to the volume fraction of each material A and B~\cite{barbero2010introduction}, while the effective thermal conductivity in the transverse direction (i.e., horizontal direction in the left figure) is given as
\begin{align}
    \kappa^\perp := \frac{1}{\frac{1}{2 \kappa^\mathrm{A}} + \frac{1}{2 \kappa^\mathrm{B}}}.
\end{align}
Here, we considered the RVE where the coordinate system is rotated $45^\circ$ clockwise from composite material in Fig.~\ref{fig:RVE_analytic_model} so that the asymptotic homogenization could not fall into the simple one-dimensional problem.
Thus, the homogenized thermal conductivity $\overline{\bm{\kappa}}^\mathrm{exact}$ of the RVE is obtained through the coordinate transformation, as follows:
\begin{align}
    \overline{\bm{\kappa}}^\mathrm{exact} &= \begin{pmatrix}
        \cos(\pi / 4) & -\sin(\pi / 4) \\
        \sin(\pi / 4) & \cos(\pi / 4)
    \end{pmatrix} \begin{pmatrix}
        \kappa^\parallel & 0 \\
        0 & \kappa^\perp
    \end{pmatrix} \begin{pmatrix}
        \cos(\pi / 4) & \sin(\pi / 4) \\
        -\sin(\pi / 4) & \cos(\pi / 4)
    \end{pmatrix} \nonumber \\
    &= \frac{1}{2} \begin{pmatrix}
        \kappa^\parallel + \kappa^\perp & \kappa^\parallel - \kappa^\perp \\
        \kappa^\parallel - \kappa^\perp & \kappa^\parallel + \kappa^\perp
    \end{pmatrix}.
\end{align}

\begin{table}[t]
    \caption{The lowest value of the maximum rank of TT to estimate the homogenized thermal conductivity tensor within the error of 0.01 for various values of the number of degrees of freedom and the threshold for truncating TTs.} \label{tbl:max_rank_analytic}
    \centering
    \begin{tabular}{cccccc}
        \hline
        & \multicolumn{5}{c}{Number of degrees of freedom} \\
        Threshold & $2^{12}$ & $2^{14}$ & $2^{16}$ & $2^{18}$ & $2^{20}$ \\
        \hline
        $10^{-4}$ & 5 & 5 & 5 & 5 & 4 \\
        $10^{-5}$ & 5 & 5 & 5 & 5 & 5 \\
        $10^{-6}$ & 5 & 5 & 5 & 5 & 5 \\
        $10^{-7}$ & 5 & 5 & 5 & 5 & 5 \\
        \hline
    \end{tabular}
\end{table}

Setting $\kappa^\mathrm{A}=1$ and $\kappa^\mathrm{B}=0.5$, we applied asymptotic homogenization for this material using the tensor-train (TT) format; the detailed homogenization method is described in Method section.
We searched for the lowest value of the maximum rank of TT to estimate the homogenized thermal conductivity tensor within the error of 0.01, that is, $\| \overline{\bm{\kappa}} - \overline{\bm{\kappa}}^\mathrm{exact} \|_2 / \| \overline{\bm{\kappa}}^\mathrm{exact} \|_2 \leq 0.01$ where $\overline{\bm{\kappa}}$ is the homogenized thermal conductivity tensor computed by the TT-based asymptotic homogenization.
We set the same maximum rank for TTs representing the material distribution and the solution of a cell problem for asymptotic homogenization.

\begin{figure}[t]
    \centering
    \includegraphics[width=0.7\textwidth]{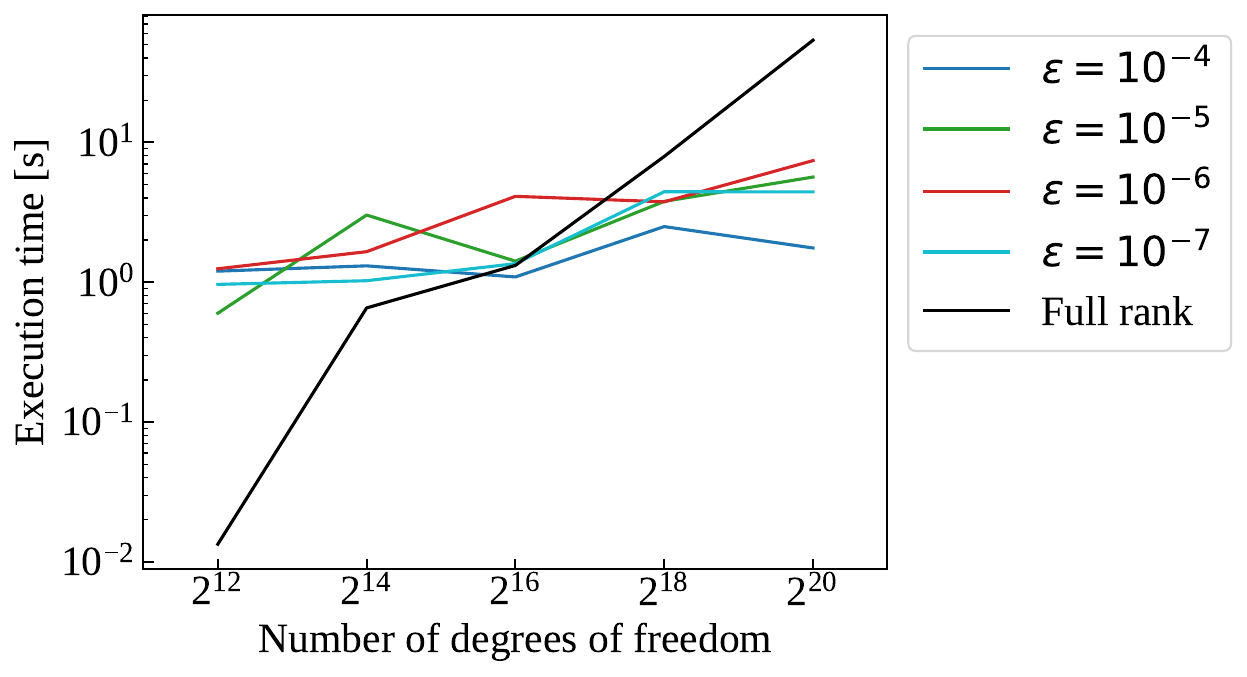}
    \caption{Computational time required for asymptotic homogenization.}
    \label{fig:result_time}
\end{figure}

Table~\ref{tbl:max_rank_analytic} lists the lowest value of the maximum rank of TT for various values of the number of degrees of freedom of the cell problem of the asymptotic homogenization and the threshold for truncating TTs.
Since the cell problem for the homogenized thermal conductivity is the problem for a scalar field, the number of degrees of freedom corresponds to the number of grid nodes, which discretizes the RVE.
How to truncate TTs is described in Supplementary information~\ref{sec:truncate_tt}, which dramatically reduces the computational cost while keeping the accuracy.
We can see in Table~\ref{tbl:max_rank_analytic} that the required number of ranks is independent of the problem size (the number of degrees of freedom), which implies the required number of ranks is inherently determined by the feature of microstructures.
Figure~\ref{fig:result_time} shows the computational times for asymptotic homogenization via the conventional finite difference method (FDM), which we call full-rank FDM, and the TT format with various values of the truncation threshold $\varepsilon$.
We observe the computational time of the proposed TT-based method is less dependent on the number of degrees of freedom compared to that of the full-rank FDM, owing to the low necessary rank.
This means that the TT format can efficiently perform asymptotic homogenization when the number of degrees of freedom is high.

\section{Generating random microstructures} \label{sec:voronoi}
In this study, we prepare the representative volume element (RVE) in a random manner using a Voronoi tessellation.
The procedure is as follows:
\begin{enumerate}
    \item Generate $N_\mathrm{point}$ in $[0, 1]^d$ uniformly at random where $d(=2, 3)$ is the number of spatial dimensions.
    \item Assign a binary 0 or 1 to each point randomly where 0 is selected at the probability of the volueme fraction $V_\mathrm{f}$.
    \item Copy the $N_\mathrm{point}$ points periodically in an enlarged domain $[-1, 2]^d$, which results in $3^d N_\mathrm{point}$ points.
    \item Perform Voronoi tessellation for $3^d N_\mathrm{point}$ points in the enlarged domain.
    \item Discretize the RVE $[0, 1]^d$ to $N^d$ grid.
    \item Assign the binary 0 or 1 to each node so that the binary could coincide with that of the nearest point, which results in the characteristic function $\chi(\bm{y})$ representing the heterogeneous material property.
\end{enumerate}

\begin{figure}[t]
    \centering
    \includegraphics[width=0.9\textwidth]{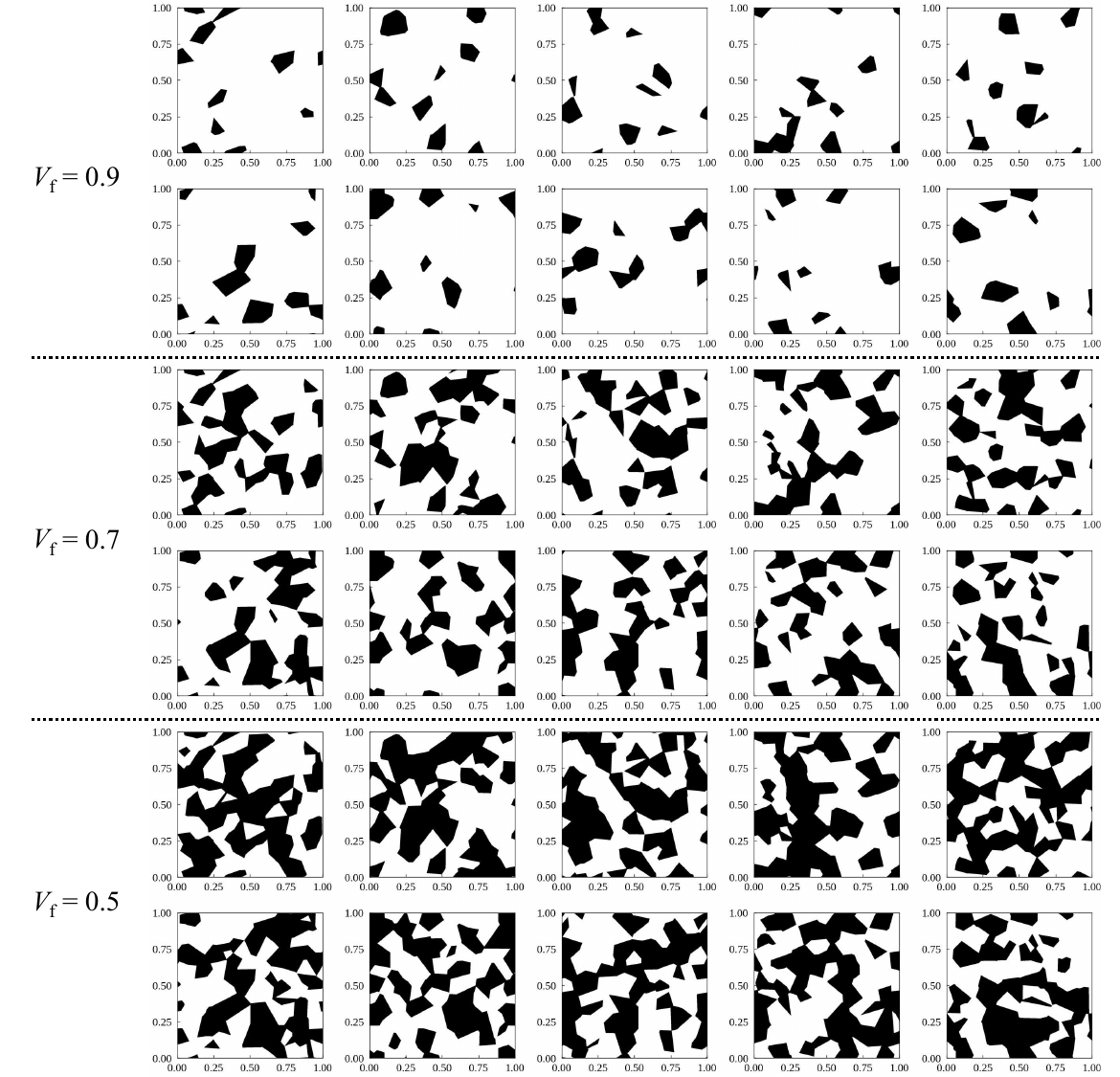}
    \caption{Randomly generated RVEs in two dimensions for various volume fractions $V_\mathrm{f}$.}
    \label{fig:random_cell_2d}
\end{figure}

\begin{figure}[t]
    \centering
    \includegraphics[width=0.9\textwidth]{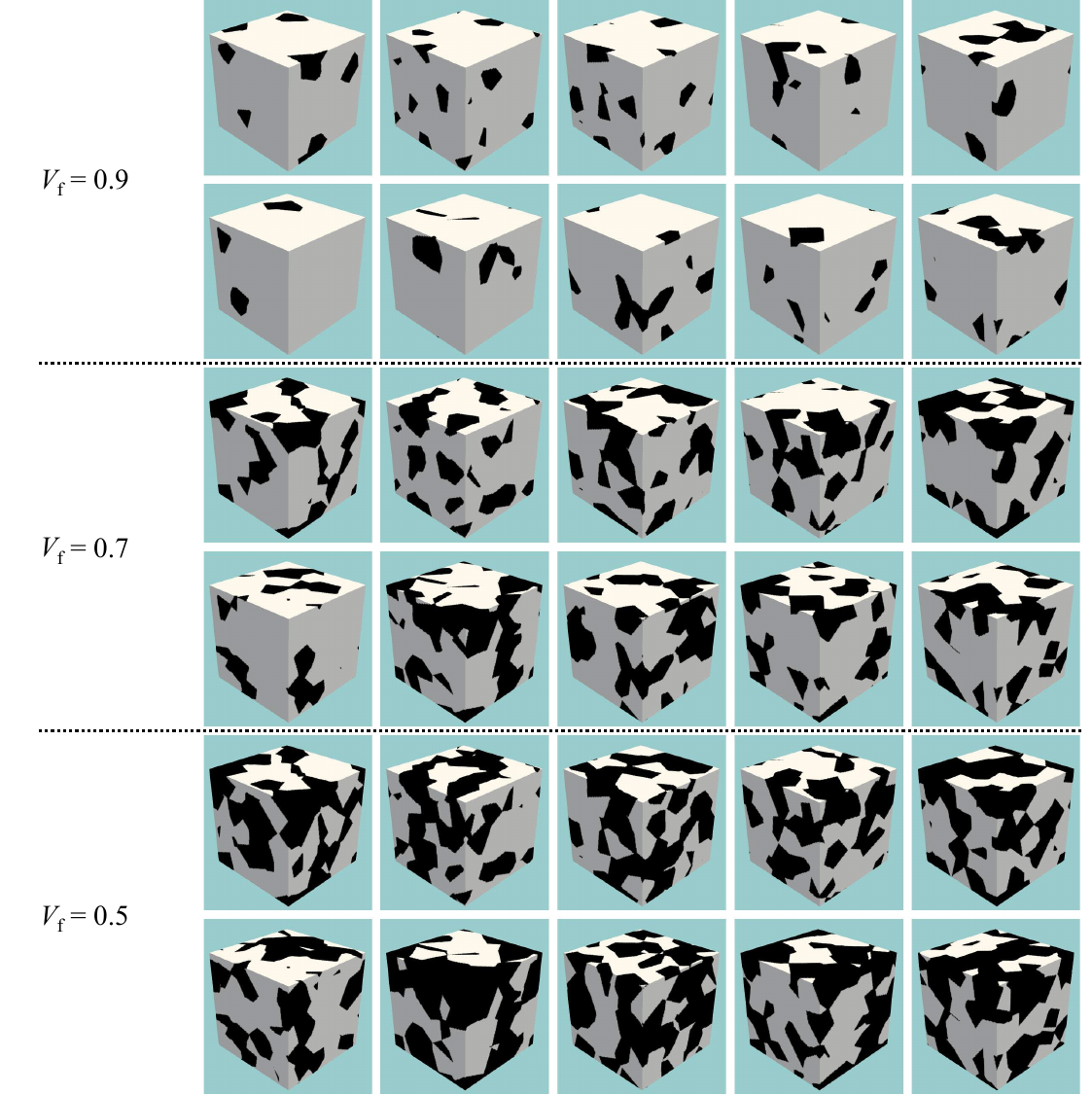}
    \caption{Randomly generated RVEs in three dimensions for various volume fractions $V_\mathrm{f}$.}
    \label{fig:random_cell_3d}
\end{figure}

The generated RVEs for volume fractions $V_\mathrm{f}=0.9, 0.7, 0.5$ are shown in Fig.~\ref{fig:random_cell_2d} for two dimensions and in Fig.~\ref{fig:random_cell_3d} for three dimensions.

\section{Addition in tensor-train formats} \label{sec:add_tt}
Let $\bm{A}$ and $\bm{B}$ be 2$m$-order tensors both in $\mathbb{R}^{d_{m-1} \times \dots \times d_0 \times d'_{m-1} \times \dots \times d'_0}$ and have the TT format respectively written as
\begin{align}
    A_{p_{m-1} \dots p_0 q_{m-1} \dots q_0} &= \bm{A}^{[m-1]}_{p_{m-1} q_{m-1}} \bm{A}^{[m-2]}_{p_{m-2} q_{m-2}} \cdots \bm{A}^{[0]}_{p_0 q_0}, \nonumber \\
    B_{p_{m-1} \dots p_0 q_{m-1} \dots q_0} &= \bm{B}^{[m-1]}_{p_{m-1} q_{m-1}} \bm{B}^{[m-2]}_{p_{m-2} q_{m-2}} \cdots \bm{B}^{[0]}_{p_0 q_0}.
\end{align}
The addition $\bm{C} := a\bm{A} + b\bm{B}$ for $a, b \in \mathbb{R}$ can be written in the TT format, as follows:
\begin{align}
    & C_{p_{m-1} \dots p_0 q_{m-1} \dots q_0} \nonumber \\
    &= a A_{p_{m-1} \dots p_0 q_{m-1} \dots q_0} + b B_{p_{m-1} \dots p_0 q_{m-1} \dots q_0} \nonumber \\
    &= a \bm{A}^{[m-1]}_{p_{m-1} q_{m-1}} \bm{A}^{[m-2]}_{p_{m-2} q_{m-2}} \cdots \bm{A}^{[0]}_{p_0 q_0} + b \bm{B}^{[m-1]}_{p_{m-1} q_{m-1}} \bm{B}^{[m-2]}_{p_{m-2} q_{m-2}} \cdots \bm{B}^{[0]}_{p_0 q_0} \nonumber \\
    &= \begin{pmatrix}
        a \bm{A}^{[m-1]}_{p_{m-1} q_{m-1}} & b \bm{B}^{[m-1]}_{p_{m-1} q_{m-1}}
    \end{pmatrix} \begin{pmatrix}
        \bm{A}^{[m-2]}_{p_{m-2} q_{m-2}} & \bm{O} \\
        \bm{O} & \bm{B}^{[m-2]}_{p_{m-2} q_{m-2}}
    \end{pmatrix} \cdots \begin{pmatrix}
        \bm{A}^{[1]}_{p_{1} q_{1}} & \bm{O} \\
        \bm{O} & \bm{B}^{[1]}_{p_{1} q_{1}}
    \end{pmatrix} \begin{pmatrix}
        \bm{A}^{[0]}_{p_0 q_0} \\
        \bm{B}^{[0]}_{p_0 q_0}
    \end{pmatrix},
\end{align}
where $\bm{O}$ represents the zero matrix with the appropriate size.
That is, for addition, it suffices to define the core of $\bm{C}$, as follows:
\begin{align}
    & \bm{C}^{[m-1]}_{p_{m-1} q_{m-1}} := \begin{pmatrix}
        a \bm{A}^{[m-1]}_{p_{m-1} q_{m-1}} & b \bm{B}^{[m-1]}_{p_{m-1} q_{m-1}}
    \end{pmatrix}, \\
    & \bm{C}^{[s]}_{p_{s} q_{s}} = \begin{pmatrix}
        \bm{A}^{[s]}_{p_{s} q_{s}} & \bm{O} \\
        \bm{O} & \bm{B}^{[s]}_{p_{s} q_{s}}
    \end{pmatrix} \text{ for } 0 < s < m-1, \\
    & \bm{C}^{[0]}_{p_0 q_0} = \begin{pmatrix}
        \bm{A}^{[0]}_{p_0 q_0} \\
        \bm{B}^{[0]}_{p_0 q_0}
    \end{pmatrix}.
\end{align}

\section{Truncation of a tensor-train format} \label{sec:truncate_tt}
Let $\bm{A} \in \mathbb{R}^{d_{m-1} \times \dots \times d_0 \times d'_{m-1} \times \dots \times d'_0}$ be a $2m$-order tensor given in the TT format as
\begin{align}
    &A_{p_{m-1} \dots p_0 q_{m-1} \dots q_0} \nonumber \\
    &=\sum_{\alpha_{m}=0}^{0} \sum_{\alpha_{m-1}=0}^{r'_{m-1}-1} \dots \sum_{\alpha_1=0}^{r'_1 - 1} \sum_{\alpha_0=0}^{0} A^{[m-1]}_{\alpha_m p_{m-1} q_{m-1} \alpha_{m-1}} \dots A^{[0]}_{\alpha_1 p_0 q_0 \alpha_0}.
\end{align}
Here, we discuss how to truncate the ranks $r'_1, \dots r'_{m-1}$ and obtain a compressed TT format with the least truncation error.
First, let us contract the cores $A^{[m-1]}_{\alpha_m p_{m-1} q_{m-1} \alpha_{m-1}}$ and $A^{[m-2]}_{\alpha_{m-1} p_{m-2} q_{m-2} \alpha_{m-2}}$ and apply the singular value decomposition to the contracted matrix, as follows:
\begin{align}
    &\sum_{\alpha_{m-1}=0}^{r'_{m-1}-1} A^{[m-1]}_{0 p_{m-1} q_{m-1} \alpha_{m-1}} A^{[m-2]}_{\alpha_{m-1} p_{m-2} q_{m-2} \alpha_{m-2}} \nonumber \\
    &= \sum_{\alpha_{m-1}=0}^{r_{m-1}-1} U_{(p_{m-1}, q_{m-1}) \alpha_{m-1}}^{[m-1]} S_{\alpha_{m-1}}^{[m-1]} V_{(q_{m-2}, q_{m-2}, \alpha_{m-2}) \alpha_{m-1}}^{[m-1]},
\end{align}
where $r_{m-1} := \min (d_{m-1} d'_{m-1}, \prod_{i=0}^{m-2} d_i d'_i)$, $U_{(p_{m-1}, q_{m-1}) \alpha_{m-1}}^{[m-1]}$ and $V_{(q_{m-2}, q_{m-2} \alpha_{m-2}) \alpha_{m-1}}^{[m-1]}$ are the unitaries and $S_{\alpha_{m-1}}^{[m-1]}$ is the singular value.
By taking the summation about $\alpha_{m-1}$ corresponding to only the first $\overline{r}_{m-1}$ largest singular values, we truncate the ranks with the truncation error
\begin{align}
    \varepsilon^2 :=&\left\| \sum_{\alpha_{m-1}=0}^{r_{m-1}-1} U_{(p_{m-1}, q_{m-1}) \alpha_{m-1}}^{[m-1]} S_{\alpha_{m-1}}^{[m-1]} V_{(q_{m-2}, q_{m-2}, \alpha_{m-2}) \alpha_{m-1}}^{[m-1]} \right. \nonumber \\
    &\quad \left. - \sum_{\alpha_{m-1}=0}^{\overline{r}_{m-1}-1} U_{(p_{m-1}, q_{m-1}) \alpha_{m-1}}^{[m-1]} S_{\alpha_{m-1}}^{[m-1]} V_{(q_{m-2}, q_{m-2}, \alpha_{m-2}) \alpha_{m-1}}^{[m-1]} \right\|_\mathrm{F}^2 \nonumber \\
    =& \sum_{\alpha_{m-1}=\overline{r}_{m-1}}^{r_{m-1}-1} \left( S_{\alpha_{m-1}}^{[m-1]} \right)^2.
\end{align}
In the similar manner, we successively truncate the ranks by contracting the neighboring cores $A^{[s]}_{\alpha_{s+1} p_{s} q_{s} \alpha_{s}}$ and $A^{[s-1]}_{\alpha_{s} p_{s-1} q_{s-1} \alpha_{s-1}}$ for an index $s$ and applying the singular value decomposition to the contracted matrix, as follows:
\begin{align}
    &\sum_{\alpha_{s}=0}^{r'_{s}-1} A^{[s]}_{\alpha_{s+1} p_{s} q_{s} \alpha_{s}} A^{[s-1]}_{\alpha_{s} p_{s-1} q_{s-1} \alpha_{s-1}} \nonumber \\
    &= \sum_{\alpha_{s}=0}^{r_{s}-1} U_{(\alpha_{s+1}, p_{s}, q_{s}) \alpha_{s}}^{[s]} S_{\alpha_{s}}^{[s]} V_{(q_{s-1}, q_{s-1}, \alpha_{s-1}) \alpha_{s}}^{[s]},
\end{align}
where $r_{s} := \min (\overline{r}_{s+1} d_{s} d'_{s}, \prod_{i=0}^{s-1} d_i d'_i)$, $U_{(\alpha_{s+1}, p_{s}, q_{s}) \alpha_{s}}^{[s]}$ and $V_{(q_{s-1}, q_{s-1}, \alpha_{s-1}) \alpha_{s}}^{[s]}$ are the unitaries and $S_{\alpha_{s}}^{[s]}$ is the singular value.
Taking the summation about $\alpha_{s}$ corresponding to only the first $\overline{r}_{s}$ largest singular values, we truncate the ranks with the truncation error of $\sum_{\alpha_{s}=\overline{r}_{s}}^{r_{s}-1} ( S_{\alpha_{s}}^{[s]})^2$ in the sense of the squared Frobenius norm.
Although the above procedure sequentially applies the singular value decomposition from the left-most core to the right-most core, it also works from the right-most core to the left-most one.